\documentclass[preprint,aps,prd,nofootinbib,floatfix]{revtex4-1}
\usepackage{latexsym,amsmath,amssymb,amsbsy,graphicx}
\usepackage{xcolor}
\usepackage{slashed} 

\allowdisplaybreaks[1]

\newcommand{\Tr}{\text{Tr}\,}

\newcommand{\nl}{\nonumber\\}

\def\BDstartaunu{{\bar B}^0 \to D^{*+} \tau^{-} {\bar\nu}_\tau}

\begin{document}

\title{Exploring new physics in the angular distribution of
 $\bar B\rightarrow D^*(\rightarrow D\pi)\tau^-(\rightarrow V^-\nu_\tau)\bar\nu_\tau$}

\author{Si-Yuan Liu${}^1$}

\author{Alakabha Datta${}^2$}
\email{datta@phy.olemiss.edu}

\author{Xian-Wei Kang${}^1$}
\email{xwkang@bnu.edu.cn}

\affiliation{${}^1$Key Laboratory of Beam Technology of Ministry of Education,  School of Physics and Astronomy, Beijing Normal University,  Beijing 100875, China\\
${}^2$Department of Physics and Astronomy, 108 Lewis Hall, University of Mississippi, Oxford, MS 38677-1848, USA}

\begin{abstract}
The decays $\bar{B}\rightarrow D^*\tau^{-}\bar{\nu}_\tau$ provide  sensitive probes of new physics in the $b\rightarrow c\tau\nu$ transition. We discuss the effects of new physics in the  decay chain $\bar{B}\rightarrow D^*(\rightarrow D\pi)\tau^{-}(\rightarrow V^-\nu_\tau)\bar{\nu}_\tau$ where $V$ is a vector meson. We first present a comprehensive analysis of the five-dimensional angular distribution for the full decay chain $\bar{B}\rightarrow D^*(\rightarrow D\pi)\tau^{-}(\rightarrow \rho^-\nu_\tau)\bar{\nu}_\tau$, exploiting the hadronic $\tau$ decay to $\rho^-\nu_\tau$ to circumvent the experimental challenge of direct $\tau$ reconstruction. The differential decay rate is expressed in terms of measurable kinematic variables
$\theta^*$, $E_\rho$, $\theta_\rho$, $\chi_\rho$, and $q^2$ -- with complete coefficient functions provided for scalar, pseudoscalar, vector, axialvector, and tensor new-physics interactions. From this distribution we construct a set of integrated observables, including the $D^*$ polarization fractions and lepton-side forward-backward and azimuthal asymmetries, which isolate distinct combinations of helicity amplitudes.
We also show that by decomposing the vector-meson energy spectrum into its
longitudinal ($L$) and transverse ($T$) helicity sectors, two additional analyzing powers ($\alpha_L$ and $\alpha_T$) related to the measurement of $\tau$ polarization
$P_\tau^{D^*}(q^2)$ are obtained. Without information on the polarization of meson $V$, the spin analyzing power of $\rho$ is much larger than that of $a_1$ with $\alpha_\rho=0.45$ and $\alpha_{a_1}=0.02$. We further perform a numerical analysis using current experimental data on $R_{D^*}$ and the $D^*$ longitudinal polarization fraction $\langle F_L^{D^*}\rangle$ to constrain the complex new-physics couplings $g_L$, $g_R$, $g_P$, and $g_T$, revealing the  correlations among them. Our formalism is also applicable to $\tau \to a_1 \nu_\tau$, where the $3\pi$ final states mainly come from the $a_1$ meson. The framework established here provides a powerful tool for disentangling new-physics scenarios with future high-statistics data.
\end{abstract}

\maketitle

\section{Introduction}

 The study of the angular distribution  for $\bar B$ decaying into $D^*$, $\tau^-$ and $\bar{\nu}_\tau$ is of great significance for the verification of the Standard Model (SM) of particle physics, as well as for exploring the new-physics (NP) effects.  The following ratios of branching fractions, $R_{D^{(*)}}$, have been measured in experiments,
\begin{equation}
    R_{D^{(*)}}=\frac{\mathcal{B}(\bar B\rightarrow D^{(*)}\tau^-\bar\nu_\tau)}{\mathcal{B}(\bar B\rightarrow D^{(*)}l^-\bar\nu_l)}.
\end{equation}
The results are compiled in Table \ref{tab:HFLAV} and show  deviations from the SM predictions.
The measured values of $R_{D^{(*)}}$, as well as the predicted values in the SM, are all based on the average values provided by the Heavy Flavor Averaging Group \cite{HFLAV}.  By considering the measured $R_D$ and $R_{D^*}$ and their correlation, the deviation from the SM reaches $3.14\sigma$.\footnote{Recent experimental updates have increased the combined deviation from the SM in $R_D$ and $R_{D^*}$ to $\sim$ 3.8$\sigma$ \cite{RDupdate}}
These data suggest that there might be NP contributions in the decay process $b\rightarrow c\tau^-\bar\nu$, see for instance, Refs.~\cite{Li:2016vvp,Li:2018rax,Hu:2018veh,Babu:2018vrl,Freytsis:2015qca,Bernlochner:2017jka} and reviews \cite{Klaver:2024xdy,Bernlochner:2021vlv}.

\begin{table}[htbp]
    \centering
    \caption{Comparison of the measured values of $R_{D^{(*)}}$ with the predicted values in the SM, provided by the Heavy Flavor Averaging Group \cite{HFLAV}. Considering the  correlation of $R_D$ and $R_{D^*}$, the combined experimental values deviates from the SM by $3.14\sigma$.}
    \label{tab:HFLAV}
   \begin{tabular*}{0.8\linewidth}{@{\extracolsep{\fill}} ccc}
        \hline\hline
         & $R_{D}$ &$R_{D^*}$ \\
        \hline
        SM & $0.296\pm0004$  &$0.254\pm0.005$ \\
        Experiment & $0.342\pm0.026$ &$0.286\pm0.012$ \\
        deviation & $2.2\sigma$ &$1.9\sigma$\\
        \hline
        \multicolumn{3}{c}{$3.14\sigma$}\\
        \hline\hline
    \end{tabular*}
\end{table}

 While  NP  can provide a  fit to the $R_{D^{(*)}}$, discriminating between competing models requires additional observables. Angular distributions are sensitive to the different coupling structures, and thus provide a more comprehensive dynamical information than the pure decay branching fractions.
 However, the complete angular reconstruction is hindered by the undetected $\nu_\tau$ from the decay of the $\tau$, making the $\tau$ momentum ${\bf p}_\tau$ experimentally inaccessible \cite{1903}. Previous studies avoided this by considering $\tau^-\rightarrow \pi^-\nu_\tau$ \cite{2005}, taking advantage of the measurable pion momentum to define a 5-dimensional angular distribution in $\bar B\rightarrow D^*(\rightarrow D\pi')\tau^-(\rightarrow \pi^-\nu_\tau)\bar\nu_\tau$.
 A step in this direction was also taken in e.g., Refs.~\cite{Nierste:2008qe,Alonso:2016gym,Bhattacharya:2024zog} where $\tau$ decays were considered for $\bar B\to D^*\tau\nu$ with general NP.

In this work, we perform a detailed analysis of the decay chain
$\bar{B}\rightarrow D^*(\rightarrow D\pi)\tau^{-}(\rightarrow V^-\nu_\tau)\bar{\nu}_\tau$, where $V$ is a vector meson. We will mainly focus on
$\bar{B}\rightarrow D^*(\rightarrow D\pi)\tau^-(\rightarrow\rho^-\nu_\tau)\bar{\nu}_\tau$ to study the possible NP in the $b\rightarrow c\tau\nu$ transition and later we will comment on the case for $V=a_1$.
Unlike previous studies that used the $\tau\rightarrow\pi\nu$ channel, we employ the hadronic decay $\tau^-\rightarrow\rho^-\nu_\tau$ to avoid the experimental difficulties associated with the direct reconstruction of the $\tau$ lepton. The calculation is more complicated than the case of the pion, due to the additional spin structure of the $\rho$ meson. The full chain produces three measurable final-state particles ($D$, $\pi$, $\rho^-$) and three intermediate states ($N^*$, $D^*$, $\tau^-$). We render a full and measurable five?dimensional angular distribution expressed in terms of the kinematic variables $\theta^*$, $E_\rho$, $\theta_\rho$, $\chi_\rho$ and $q^2$, in the presence of scalar, pseudoscalar, vector, axialvector and tensor NP interactions. By performing a full angular analysis, different  NP contributions can be discriminated.

Based on this full angular distribution, we construct a set of integrated observables, including the $D^*$ polarization fractions, the forward?backward asymmetry $\eta_1(E_\rho,q^2)$, and the T-odd correlation $\eta_2(E_\rho,q^2)$ arising from the $\sin{2\chi_\rho}$ terms. Furthermore, using current experimental data on $R_{D^*}$ and the $D^*$ polarization, we perform a numerical constraint analysis on the complex couplings $g_{L}$, $g_{R}$, $g_{P}$, $g_{T}$.

The prerequisite for using this method is that the momentum of the decaying $B$ must be known. Therefore, the technology described in this paper is more suitable for experiments in $e^+e^-$ machines, such as Belle II.

The remainder of this paper is organized as follows. In Sec.~\ref{Angular Distribution}, we derive the five?dimensional angular distribution, with the explicit coefficient functions for scalar, pseudoscalar, vector, axialvector and tensor NP interactions. In Sec.~\ref{Integrated Observables} we construct several integrated observables, which can be  measured in experiments. In Sec.~\ref{pp}, we show how the spin analyzing power is obtained from the helicity-separated $\rho$ energy spectrum, related to the extraction of the $\tau$ polarization $P_\tau^{D^*}(q^2)$. In Sec.~\ref{sec:numerical} we give numerical constraints on NP couplings from $R_{D^*}$ and the $D^*$ polarization. We conclude in Sec~\ref{Conclusion}. Appendices contain some useful details of our calculations.

\section{Angular Distribution}
\label{Angular Distribution}
 In this section, we provide the result for the full angular distribution of $\bar B\rightarrow D^* (\rightarrow D\pi)\tau(\rightarrow \rho \nu_\tau)\bar\nu_\tau$ process.

\subsection{Decay kinematics}
\label{Angular distribution structure}
\begin{figure}
    \centering
    \includegraphics[width=0.75\linewidth]{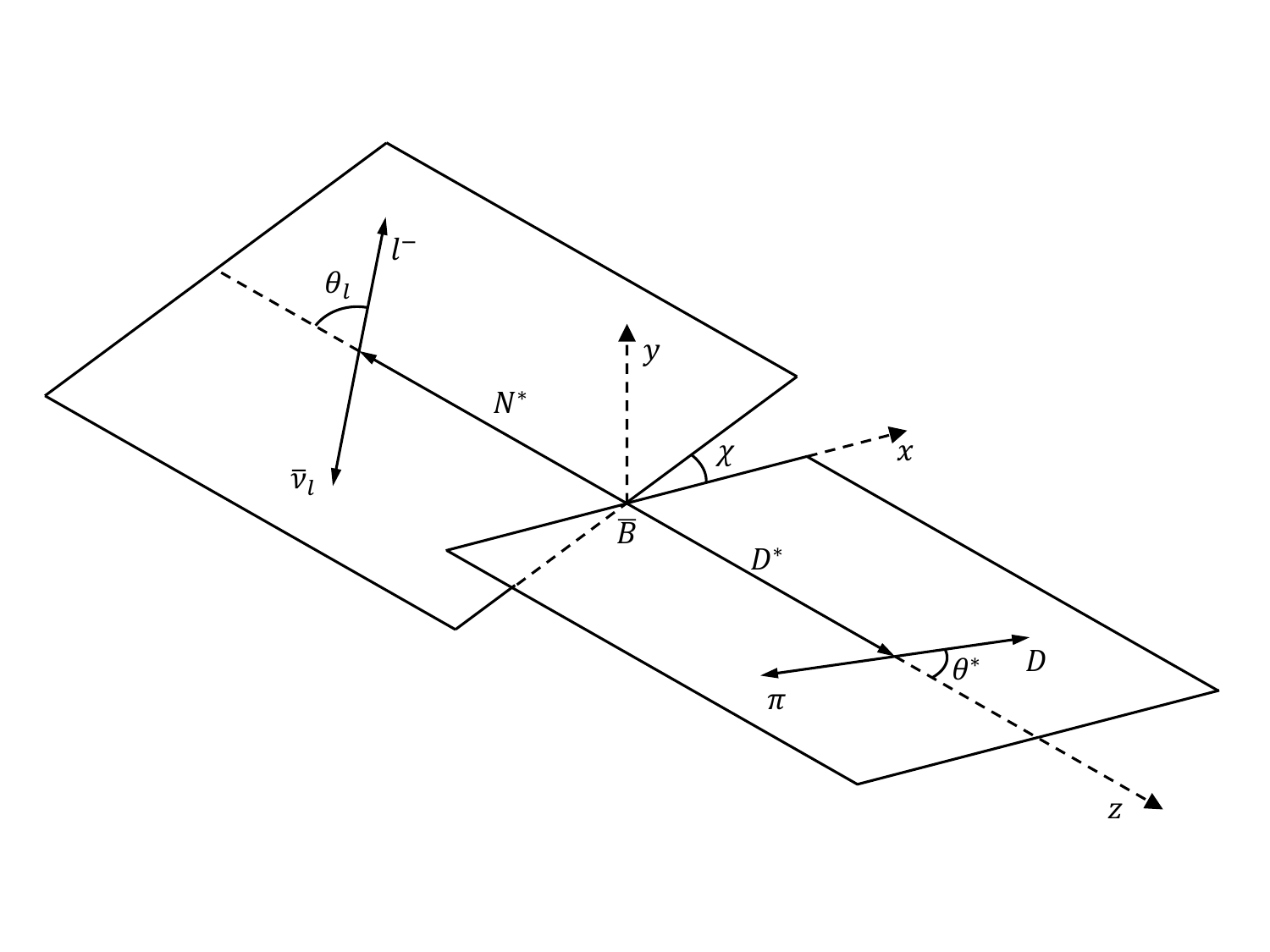}
    \caption{Definition of the angles in the $\bar B\rightarrow D^*(\rightarrow D\pi)l^-\bar\nu_l$ distribution.}
    \label{Definition_l}
\end{figure}

 First, consider the angular distribution of the decay $\bar B\rightarrow D^* (\rightarrow D\pi)\tau\bar\nu_\tau$. Suppose that the particle that mediates the weak interaction is $N$, the reaction would be written as: $\bar B\rightarrow D^* (\rightarrow D\pi) N^*(\rightarrow l^-\bar\nu_l)$, where $N=S, P, V, A, T$ represent scalar, pseudoscalar, vector, axial vector and tensor interactions, respectively. For the case where $l=e,\mu$, we assume that no NP is present and only the SM is involved, so $N$ is just the $W$ boson in the SM. But for $l=\tau$, all couplings, including NP ones, are allowed. The decay amplitude, as a function of the final-state momenta, will be represented by the three angles defined in Fig.~\ref{Definition_l}, which is the angular distribution. The four-body final state process, $\bar B\rightarrow D^* (\rightarrow D\pi)l\nu$ has been studied extensively, see, for example, Refs.~\cite{Kang:2013jaa,Duraisamy:2013pia,Duraisamy:2014sna}.

In Fig.~\ref{Definition_l}, we define a right-handed coordinate system by aligning the $z$ axis with the $D^*$ momentum, taking the $y$ axis parallel to ${\bf p}_{D^*}\times{\bf p}_{D}$, and setting the $x$ axis by $\hat{\bf{x}}=\hat{\bf{y}}\times\hat{\bf{z}}$.
The angle $\theta^*$ represents the three-momentum polar angle of the $D$ meson in the rest frame of the $D^*$ meson. The $\theta_l$ and $\chi$ represent the polar and azimuthal angle of the leptons in the rest frame of their parent particles.

Consider now the case where the lepton $l$ is $\tau$. In experiments, ${\bf p}_\tau$ cannot be detected directly, but only its decay products can be observed. That is, $\tau$ will be reconstructed through observable physical quantities such as the energy and momentum of its decay products. A measurable angular distribution of the simplest decay $\tau\rightarrow \pi \nu_\tau$ has already been presented in Ref.~\cite{1903}. In this article, we choose another decay channel $\tau\rightarrow \rho \nu_\tau$ and calculate the measurable angular distribution. Furthermore, we assume that NP exists in the decay $\bar B\rightarrow D^*N^*$ but not in the decays of the $\tau$.

 The complete decay chain $\bar B\rightarrow D^* (\rightarrow D\pi) N^*[\rightarrow\tau(\rightarrow \rho \nu_\tau)\bar\nu_\tau]$, consists of five final-state particles: $D$, $\pi$, $\bar\nu_\tau$, $\rho$ and $\nu_\tau$, and three intermediate-state particles $D^*$, $N^*$ and $\tau$. Therefore, its phase space depends on eight independent parameters: five relevant angles and three squared values of the invariant masses corresponding to the intermediate particles. Using the on-shell approximation for $D^*$ and $\tau$, two of the three invariant mass squares are given by $m_{D^*}$ and $m_\tau$. Therefore, this decay depends on six independent parameters: five helicity angles and the squared momentum transfer $q^2$.

Usually, five helicity angles should be adopted: (i) the polar angle $\theta^*$ of the $D$ meson in the $D^*$ rest frame, (ii) the polar angle $\theta_\tau$ and the azimuthal angle $\chi_\tau$ of the $\tau$ in the $N^*$ rest frame, (iii) the polar angle $\theta_\rho$ and the azimuthal angle $\chi_\rho$ of the $\rho$ in the $\tau$ rest frame. However, because the helicity angles of $\tau$ cannot be measured, it is necessary to select other measurable angles. An effective way is to express the kinematics of the $\tau$ decay products in the $N^*$ rest frame.

\begin{figure}
    \centering
    \includegraphics[width=0.75\linewidth]{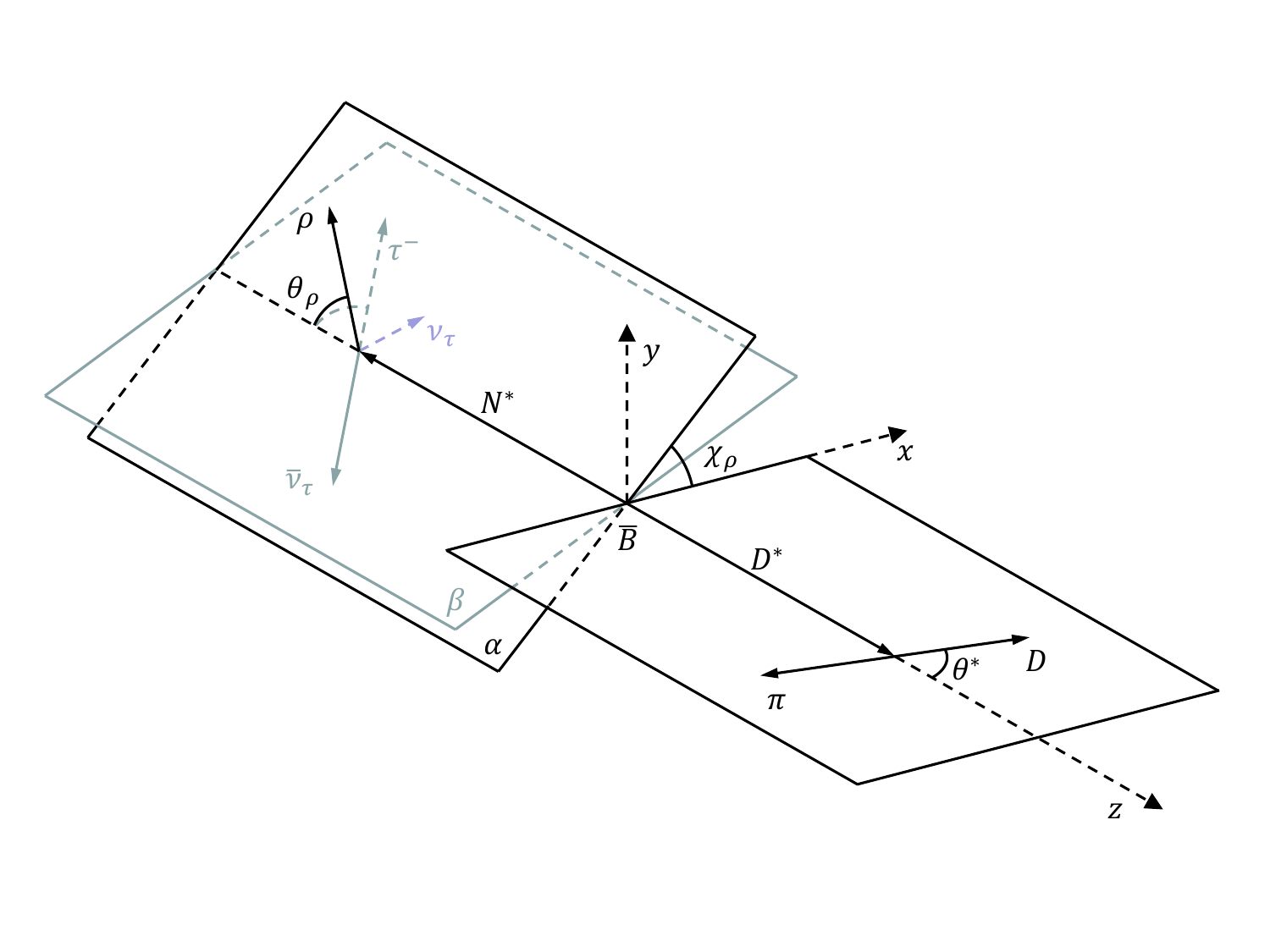}
    \caption{Definition of the angles in the $\bar B\rightarrow D^*(\rightarrow D\pi)\tau^-(\rightarrow \rho^-\nu_\tau)\bar\nu_\tau$ distribution.}
    \label{Definition_r}
\end{figure}

Fig.~\ref{Definition_r} shows the kinematics of the decay process $\bar{B}\to D^*(\to D\pi)\tau^-(\to\rho^-\nu_\tau)\bar{\nu}_\tau$. In Fig.~\ref{Definition_r}, the definitions of the $xyz$ axes are the same as those in the Fig.~\ref{Definition_l}. The plane I, composed of dark-colored line segments, is the plane where both $N^*$ and $\rho$ exist. The plane II, composed of light-colored line segments, is the plane where both $N^*$ and $\tau^-$ exist. Note that $\nu_\tau$ may not be within these two planes.
Here we adopt the measurable helicity angles:
 (i) the polar angle $\theta^*$ of the $D$ meson in the $D^*$ rest frame, (ii) the polar angle $\theta_\rho$ and the azimuthal angle $\chi_\rho$ of the $\rho$ in the $N^*$ rest frame. (iii) the polar angle $\theta_{\tau\rho}$ and azimuthal angle $\chi_{\tau\rho}$ are formed by the momentum of the $\tau$ and $\rho$ in the $N^*$ rest frame, where $\chi_{\tau\rho}$ cannot be determined and will be finally integrated over.

Based on the above definitions, we can express the differential decay rate of the entire process as
\begin{equation}
\frac{d^5\Gamma}{d\cos\theta^*dE_\rho d\cos\theta_\rho d\chi_\rho dq^2 }=\frac{1}{2^{17}\pi^{9}}\frac{1}{\sqrt{q^2}}\frac{|{\bf p}_{D^*}||{\bf p}_D|}{m_B^2m_{D^*}}\int |\mathcal{M}|^2 dp_{D^*}^2 dp_\tau^2d\chi_{\tau\rho},
\end{equation}
and the relevant kinematic variables are
\begin{align}
    |{\bf p}_{D^*}|&=\frac{\sqrt{(m_B^2-m_{D^*}^2-q^2)^2-4m_{D^*}^2q^2}}{2m_B},\nl
    |{\bf p}_D|&=\frac{\sqrt{(m_{D^*}^2-m_{D}^2-m_{\pi}^2)^2-4m_{D}^2m_\pi^2}}{2m_D},\nl
    E_\tau&=\frac{q^2+m_\tau^2}{2\sqrt{q^2}}, \quad |{\bf p}_{\tau}|=\frac{q^2-m_\tau^2}{2\sqrt{q^2}},\nl
    \cos\theta_{\tau\rho}&=\frac{2E_\tau E_\rho-m_\tau^2-m_\rho^2}{2|{\bf p}_\tau||{\bf p}_\rho|}.
\end{align}
 For the calculation of phase space of this reaction, we  refer to Appendix A for details.

\subsection{Momenta and polarization vectors in the $W^*$ rest frame}
 In the $B$-meson rest frame, the polarization vectors of the $D^*$ meson and the $W^*$ meson are
\begin{eqnarray}\label{eq:polarizationD*}
    &&\epsilon_{D^*}^{\mu}(\pm)=(0,1,\pm i,0)/\sqrt{2},\quad \epsilon_{D^*}^{\mu}(0)=(k_z,0,0,k_0)/m_{D^*},\\
    &&\epsilon_{W^*}^{\mu}(\pm)=(0,1,\mp i,0)/\sqrt{2},\quad \epsilon_{W^*}^{\mu}(0)=-(q_z,0,0,q_0)/\sqrt{q^2},\quad \epsilon_{W^*}^{\mu}(t)=q^\mu/\sqrt{q^2}.
\end{eqnarray}
 where $k^\mu=(k_0,0,0,k_z)$ and $q^\mu=(q_0,0,0,q_z)$ are the four momenta of the $D^*$ and the $W^*$ in the $B$-meson rest frame. The polarization vectors of the off-shell $W^*$ satisfy the following orthonormality and completeness relations:
    \begin{align}
        \epsilon^{*\mu}_{W^*}(m)\epsilon_{W^*\mu}(m')&=g_{mm'},\\
        \sum_{m,m'}\epsilon^{*\mu}_{W^*}(m)\epsilon_{W^*}^\nu(m')g_{mm'}&=g^{\mu\nu},
    \end{align}
where $m, m'$ can take $t, \pm1, 0$, and $g_{mm'}$ is the diagonal matrix with $g_{tt}=1, g_{11}=g_{-1-1}=g_{00}=-1$.
The polarizations of the $D^*$ satisfy these relations:
    \begin{eqnarray}
    \label{eq:D*pola_relation}
        \epsilon^{*\mu}_{D^*}(m)\epsilon_{D^*\mu}(m')&=&-\delta_{mm'},\\
        \sum_{m,m'}\epsilon^{*\mu}_{D^*}(m)\epsilon_{D^*}^\nu(m')g_{mm'}&=&-g^{\mu\nu}+\frac{k^\mu k^\nu}{m_{D^*}^2}.
    \label{eq:D*pola_relation2}
    \end{eqnarray}

With the $\rho$ momentum aligned along the $z$-axis, its polarization vector takes the form
\begin{equation}
    \epsilon^{\mu}_{\rho}(\pm)=(0,1,\pm i,0)/\sqrt{2},\quad \epsilon^{\mu}_{\rho}(0)=(|{\bf p}_{\rho}|,0,0,E_\rho)/m_\rho,
\end{equation}
where $p_\rho=(E_\rho,{\bf p}_\rho)$ are the four momenta of the $\rho$ in the $N^*$ rest frame. However, to calculate the angular distribution, we must rotate the polarization vector of $\rho$ to the $xyz$ coordinate system defined in Fig.~\ref{Definition_r}. The polarization vector of the $\rho$ in the $W^*$ rest frame can be obtained by a Wigner rotation as follows:
\begin{equation}
    \epsilon(m')=\sum_{m=0,\pm1}D^1_{m'm}(\alpha,\beta,\gamma)\epsilon(m).
\end{equation}
with
\begin{equation}
 D^1(\alpha,\beta,\gamma)=
    \begin{pmatrix}
 e^{-i(\alpha+\gamma)}\frac{1+\cos\beta}{2} & -e^{-i\alpha}\frac{\sin\beta}{\sqrt{2}} & e^{-i(\alpha-\gamma)}\frac{1-\cos\beta}{2} \\
 e^{-i\gamma}\frac{\sin\beta}{\sqrt{2}} & \cos\beta & -e^{i\gamma}\frac{\sin\beta}{\sqrt{2}} \\
 e^{i(\alpha-\gamma)}\frac{1-\cos\beta}{2}&
 e^{i\alpha}\frac{\sin\beta}{\sqrt{2}}& e^{i(\alpha+\gamma)}\frac{1+\cos\beta}{2}
\end{pmatrix},
\end{equation}
where $\alpha$, $\beta$, and $\gamma$ denote Euler angles that uniquely parameterize a three-dimensional rotation. Here, \( \alpha \) is set to be \( \chi_\rho \), \( \beta \) is set to \( \pi - \theta_\rho \), and \( \gamma \) to \( -\chi_\rho \). The choice of the angle $\gamma$ is purely a convention \cite{Richman} and does not affect the physics. Then we get
\begin{equation}
    \begin{split}
        \epsilon_{\rho}^{\mu}(+)&=\frac{\sqrt{2}}{2}e^{i\chi_\rho}\left(0,\cos\theta_\rho\cos\chi_\rho+i \sin\chi_\rho,\cos\theta_\rho\sin\chi_\rho-i\cos\chi_\rho,\sin\theta_\rho\right),\\
        \epsilon_{\rho}^\mu(0)&=\frac{1}{m_\rho}\left(|{\bf p}_\rho|,E_\rho\cos\chi_\rho\sin\theta_\rho,E_\rho\sin\theta_\rho\sin\chi_\rho,-E_\rho\cos\theta_\rho\right),\\
        \epsilon_{\rho}^\mu(-)&=\frac{\sqrt{2}}{2}e^{-i\chi_\rho}\left(0,-\cos\theta_\rho\cos\chi_\rho+i\sin\chi_\rho,-\cos\theta_\rho\sin\chi_\rho-i\cos\chi_\rho,-\sin\theta_\rho\right),
    \end{split}
\end{equation}
which also satisfy the relations as in Eqs.~\eqref{eq:D*pola_relation} and \eqref{eq:D*pola_relation2}.

Similarly, we should also express the $\rho$ and $\tau$ momentum in the $W^*$ rest frame. Referring to Fig.~\ref{Definition_r}, we first rotate the momentum of the $\rho$ by an angle $\pi-\theta_\rho$ around the $y$-axis and then by an angle $\chi_\rho$ around the $z$-axis \footnote{Here the angles are already defined in the $W^*$ rest frame, and only a rotation is involved. This is unlike the case in Refs.~\cite{Kang:2013jaa,Cabibbo:1965zzb}, where a Lorentz boost is needed.}. As a result,  $p_\rho$ and $p_\tau$ in the $W^*$ rest frame are
\begin{align}
p_\rho&=({E_\rho},|\bf{p}_\rho|\cos\chi_\rho\sin\theta_\rho,|\bf{p}_\rho|\sin\chi_\rho\sin\theta_\rho,-|\bf{p}_\rho|\cos\theta_\rho),\nl
p_\tau&=\Big({E_\tau},|\bf{p}_\tau|\big[\cos\theta_{\tau\rho}\cos\chi_\rho\sin\theta_\rho-\sin\theta_{\tau\rho}(\cos\theta_\rho\cos\chi_\rho\cos\chi_{\tau\rho}+\sin\chi_\rho\sin\chi_{\tau\rho})\big],\nl
&\quad\:|\bf{p}_\tau|\big[\cos\theta_{\tau\rho}\sin\chi_\rho\sin\theta_\rho+\sin\theta_{\tau\rho}(-\cos\theta_\rho\sin\chi_\rho\cos\chi_{\tau\rho}+\cos\chi_\rho\sin\chi_{\tau\rho})\big],\nl
&\quad\:-|\bf{p}_\tau|(\cos\theta_{\tau\rho}\cos\theta_\rho+\sin\theta_{\tau\rho}\sin\theta_\rho\cos\chi_{\tau\rho})\Big).
\end{align}
The same result can also be achieved by the above Wigner $D$ function by setting the Euler angle $\gamma=0$.

\subsection{Amplitude modulus squared $|\mathcal{M}|^2$}
In this subsection, we calculate the modulus squared of the amplitude,  $|\mathcal{M}|^2$. In the presence of NP,
we consider the effective Hamiltonian:
\begin{eqnarray}
\mathcal{H}_{\text{eff}}&=&\frac{G_{\text{F}} V_{cb}}{\sqrt{2}}\Big\{\big[(1+g_{L})\bar c \gamma_\alpha (1-\gamma_5) b+g_{R}\bar c \gamma_\alpha (1+\gamma_5) b\big] \bar\tau\gamma^\alpha(1-\gamma_5)\nu_\tau \nonumber\\
&&+\big[g_{S}\bar c b+g_{P}\bar c \gamma_5 b\big]\bar\tau(1-\gamma_5)\nu_\tau +g_{T}\bar c \sigma^{\alpha\beta} (1-\gamma_5)b\bar\tau \sigma_{\alpha\beta}(1-\gamma_5)\nu_\tau\Big\}+h.c.,
\end{eqnarray}
where $h.c.$ denotes the hermitian conjugate, $S, P, V, A, T$ represent scalar, pseudoscalar, vector, axialvector, and tensor interactions, respectively; $G_{\text{F}}$ is the Fermi constant, and $V_{cb}$ is the CKM matrix element. The  SM corresponds to $g_{S}=g_{P}=g_{L}=g_{R}=g_{T}=0$.
Note also that the coupling $g_S$  does not contribute to the decay $\bar{B}\to D^*\tau^-\bar{\nu}_{\tau}$.


The  amplitude for the decay process $\bar B\rightarrow D^*(\to D\pi )W^*(\rightarrow\tau^-\bar\nu_\tau)$ is \cite{1903}
\begin{align}
\mathcal{M}^{\operatorname{SM+NP}}\propto
&\sum_{m,m'=\pm,0}\epsilon^\nu_{D^*}(m)(p_D)_\nu g_{mm'}\epsilon_{D^*}^{*\mu}(m')M_\mu^{SP}(\bar u_\tau (1-\gamma_5) v_{\bar\nu_\tau})\nl
&+\sum_{m,m'=\pm,0}\epsilon_{D^*}^\alpha(m)(p_D)_\alpha g_{mm'} \epsilon^{*\mu}_{D^*}(m')M_{\mu\nu}^{VA} \times \sum_{n,n'=t,\pm 0}\epsilon^{*\nu}_{VA}(n')g_{n'n}\epsilon_{VA}^\xi(n)(\bar u_{\tau} \gamma_{\xi} (1-\gamma_5) v_{\bar\nu_\tau})\nl
&+\sum_{m,m'=\pm,0}\epsilon_{D^*}^\alpha(m)(p_D)_\alpha g_{mm'}\epsilon_{D^*}^{*\mu}(m')M_{\mu,\nu\sigma}^T\times\sum_{n,n'=t,\pm,0}\epsilon_T^{*\nu}(n')g_{n'n}\epsilon_T^{\xi}(n)\nl
&\qquad\times\sum_{p,p'=t,\pm,0}\epsilon_{T}^{*\sigma}(p')g_{p'p}\epsilon_{T}^{\zeta}(p)(\bar u_\tau \sigma_{\xi\zeta} (1-\gamma_5) v_{\bar\nu_\tau}),
\end{align}
where $M^{SP}_\mu$, $M_{\mu\nu}^{VA}$, $M_{\mu,\nu\sigma}^T$ are the transition matrix elements describing $B\to D^*N^*$ with the corresponding interactions and the details can be found in Ref.~\cite{1903}. The decay amplitude of $\tau^-\to\rho^-\nu_\tau$ is given by
\begin{equation}
\mathcal{M}_{\tau }(m)=\frac{G_{\text{F}} V_{ud}}{\sqrt{2}}f_{\rho }m_{\rho }
[\bar u_{\nu_\tau}\gamma_\mu(1-\gamma_5)u_\tau]
\epsilon _{\rho }^{*\mu }(m),
\end{equation}
where $f_\rho$ is the decay constant of the $\rho$ meson, and will be fixed by its decay rate.

Therefore, the total decay amplitude of the five-body decay process $\bar B^0\rightarrow D^{*+}(\rightarrow D^0\pi^+)\tau^-(\rightarrow\rho^- \nu_{\tau})\bar\nu_\tau$ is
\begin{equation}
    \mathcal{M}(k)\propto \sum_\text{spins} \mathcal{M}_\tau(k) \mathcal{M}^{\operatorname{SM+NP}},
\end{equation}
with $k$ denoting the polarization of the $\rho$ meson and $k=0,\pm$.
After simplification, we obtain
\begin{align}
    \mathcal{M}(k)\propto
     -\sum_{m=\pm,0}&\mathcal{H}_{D^*}(m)\Big\{\mathcal{M}_{SP}(m)\mathcal{L}_{SP}(k)+\sum_{n=t,\pm,0} g_{nn}\mathcal{M}_{VA}(m,n)\mathcal{L}_{VA}(n,k)\nl    +&\sum_{n,p=t,\pm,0}g_{nn}g_{pp}\mathcal{M}_{T}(m,n,p)\mathcal{L}_T(n,p,k)\Big\},
\end{align}
where
\begin{align}
    \mathcal{H}_{D^*}(m)&=\epsilon_{D^*}(m)\cdot p_D,\nl
    \mathcal{M}_{SP}(m) &=\epsilon_{D^*}^{*\mu}(m)M_{\mu}^{SP},\nl
    \mathcal{M}_{VA}(m,n)&=\epsilon_{D^*}^{*\mu}(m)M_{\mu\nu}^{VA}\epsilon^{*\nu}_{VA}(n),\nl
    \mathcal{M}_{T}(m,n,p)&=\epsilon_{D^*}^{*\mu}(m)M^{T}_{\mu,\nu\sigma}\epsilon^{*\nu}_T(n)\epsilon^{*\sigma}_T(p),\nl
    \mathcal{L}_{SP}(k)&=(m_\tau \bar u_{\nu_\tau}\gamma_{\mu}(1-\gamma_5) v_{\bar\nu_\tau})\epsilon_\rho^{*\mu}(k),\nl
    \mathcal{L}_{VA}(n,k)&=\epsilon^{\mu}_{VA}(n)(\bar u_{\nu_\tau}\gamma_{\nu}  \slashed{p}_{\tau} \gamma_\mu (1-\gamma_5) v_{\bar\nu_\tau})\epsilon_{\rho}^{*\nu}(k),\nl
    \mathcal{L}_{T}(n,p,k)&=\epsilon^\mu_{T}(n)\epsilon^\nu_T(p)(m_\tau\bar u_{\nu_\tau}\gamma_\xi\sigma_{\mu\nu}(1-\gamma_5) v_{\bar\nu_\tau})\epsilon^{*\xi}_\rho(k),
\end{align}
and the minus sign in front comes form $g_{mm}=-1$ with $m=\pm,0$.

Only some specific configurations of helicity components for the produced states are allowed due to angular momentum conservation. The spin of the $B$ meson is 0 and so for the $SP$ interaction, $N^*$ has spin 0, and thus the helicity component of $D^*$ can only be 0. For the $VA$ interaction, there are four allowed helicity combinations of $N^*$ and $D^*$. For the $T$  interaction, there are six permissible helicity configurations for $N^*$ and $D^*$. As a result, the non-vanishing helicity amplitudes are
    \begin{align}
        \mathcal{M}_{SP}(0)&=\mathcal{A}_{SP},\nl
        \mathcal{M}_{VA}(+,+)&=\mathcal{A}_{+},\nl
        \mathcal{M}_{VA}(-,-)&=\mathcal{A}_{-},\nl
        \mathcal{M}_{VA}(0,0)&=\mathcal{A}_0,\nl
        \mathcal{M}_{VA}(0,t)&=\mathcal{A}_t,\nl
        \mathcal{M}_{T}(+,+,0)&=\mathcal{M}_{T}(+,+,t)=\mathcal{A}_{+,T},\nl
        \mathcal{M}_{T}(0,-,+)&=\mathcal{M}_{T}(0,0,t)=\mathcal{A}_{0,T},\nl
        \mathcal{M}_{T}(-,0,-)&=\mathcal{M}_{T}(-,-,t)=\mathcal{A}_{-,T}.
    \end{align}

These non-vanishing helicity amplitudes are related to the  transition form factors ~\cite{1903}:
    \begin{align}
    \label{eq:Ai}
    \mathcal{A}_{SP}=&-g_{P}\frac{\sqrt{\lambda(m_B^2,m_{D^*}^2,q^2)}}{m_b+m_c}A_0(q^2),\nl
    \mathcal{A}_0=&-\frac{(1+g_{L}-g_{R})(m_B+m_{D^*})}{2m_{D^*}\sqrt{q^2}}\Big( (m_B^2-m_{D^*}^2-q^2)A_1(q^2)+\frac{\lambda(m_B^2,m_{D^*}^2,q^2)}{(m_B+m_{D^*})^2}A_2(q^2) \Big),\nl
    \mathcal{A}_t=&-(1+g_{L}-g_{R})\frac{\sqrt{\lambda(m_B^2,m_{D^*}^2,q^2)}}{\sqrt{q^2}}A_0(q^2),\nl
    \mathcal{A}_\pm=&(1+g_{L}-g_{R})(m_B+m_{D^*})A_1(q^2)\mp(1+g_{L}+g_{R})\frac{\sqrt{\lambda(m_B^2,m_{D^*}^2,q^2)}}{m_B+m_{D^*}}V(q^2),\nl
    \mathcal{A}_{0,T}=&-\frac{g_{T}}{2m_{D^*}}\Big( (m_B^2+3m_{D^*}^2-q^2)T_2(q^2)-\frac{\lambda(m_B^2,m_{D^*}^2,q^2)T_3(q^2)}{m_B^2-m_{D^*}^2}\Big),\nl
    \mathcal{A}_{\pm,T}=&g_{T}\frac{\pm\sqrt{\lambda(m_B^2,m_{D^*},q^2)}T_1(q^2)+(m_B^2-m_{D^*}^2)T_2(q^2)}{\sqrt{q^2}},
    \end{align}
with $\lambda(a,b,c)=a^2+b^2+c^2-2ab-2bc-2ca$. Furthermore, we can choose the transversity basis:
\begin{align}
        \mathcal{A}_\parallel&=\frac{\mathcal{A}_++A_-}{\sqrt{2}},\nl
        \mathcal{A}_\perp&=\frac{\mathcal{A}_+-A_-}{\sqrt{2}},\nl
        \mathcal{A}_{\parallel,T}&=\frac{\mathcal{A}_{+,T}+A_{-,T}}{\sqrt{2}},\nl
        \mathcal{A}_{\perp,T}&=\frac{\mathcal{A}_{+,T}-A_{-,T}}{\sqrt{2}}.
\end{align}

The next step is to calculate the modulus squared of the total decay amplitude,
\begin{align}
        |\mathcal{M}|^2&\propto\sum_\text{spins}\Big|\sum_{k=\pm,0}\mathcal{M}(k)\Big|^2\nl
        &\propto\sum_\text{spins}\Big(\sum_{k_1,m_1=\pm,0}\mathcal{H}_{D^*}(m_1)\{\mathcal{M}_{SP}(m_1)\mathcal{L}_{SP}(k_1)+\sum_{n_1=t,\pm,0} g_{n_1n_1}\mathcal{M}_{VA}(m_1,n_1)\mathcal{L}_{VA}(n_1,k_1)\nl
      &\quad+\sum_{n_1,p_1=t,\pm,0}g_{n_1n_1}g_{p_1p_1}\mathcal{M}_T(m_1,n_1,p_1)\mathcal{L}_T(n_1,p_1,k_1)\}\Big)\nl
    &\times\Big(\sum_{k_2,m_2=\pm,0}\mathcal{H}_{D^*}(m_2)\{\mathcal{M}_{SP}(m_2)\mathcal{L}_{SP}(k_2)+\sum_{n_2=t,\pm,0} g_{n_2n_2}\mathcal{M}_{VA}(m_2,n_2)\mathcal{L}_{VA}(n_2,k_2)\nl
    &\quad+\sum_{n_2,p_2=t,\pm,0}g_{n_2n_2}g_{p_2p_2}\mathcal{M}_{T}(m_2,n_2,p_2)\mathcal{L}_T(n_2,p_2,k_2)\}\Big)^\dagger\nl
    &=\mathcal{M}_{SP-SP}+\mathcal{M}_{VA-VA}+\mathcal{M}_{T-T}\nl
    &\quad+\mathcal{M}_{SP-VA}+\mathcal{M}_{SP-T}+\mathcal{M}_{VA-SP}+\mathcal{M}_{VA-T}+\mathcal{M}_{T-SP}+\mathcal{M}_{T-VA},
\end{align}
with
\begin{align}
 \mathcal{M}_{SP-SP}&=\sum \mathcal{H}_{D^*}(m_1)\mathcal{H}_{D^*}^*(m_2)\mathcal{M}_{SP}(m_1)\mathcal{M}_{SP}^*(m_2)\nl
        &\times m_{\tau}^2 \epsilon_{\rho}^{*\xi}(k_1)\epsilon_{\rho}^{\zeta}(k_2)\Tr[\slashed{p}_{\nu_\tau}\gamma_{\xi}(1-\gamma_5)\slashed{p}_{\bar\nu_\tau}\gamma_{\zeta}(1-\gamma_5)],
       \\[2ex]
        \mathcal{M}_{VA-VA}&=\sum g_{n_1n_1}g_{n_2n_2}\mathcal{H}_{D^*}(m_1)\mathcal{H}_{D^*}^*(m_2)\mathcal{M}_{VA}(m_1,n_1)\mathcal{M}_{VA}^*(m_2,n_2)\nl
        &\times\epsilon_{VA}^{\mu}(n_1)\epsilon_{VA}^{*\nu}(n_2)\epsilon^{*\alpha}_{\rho}(k_1)\epsilon^{\beta}_{\rho}(k_2)
        \Tr[\slashed{p}_{\nu_\tau}\gamma_{\alpha}\slashed{p}_{\tau}\gamma_{\mu}(1-\gamma_5) \slashed{p}_{\bar\nu_\tau}\gamma_{\nu}\slashed{p}_\tau\gamma_{\beta}(1-\gamma_5)],
        \\[2ex]
        \mathcal{M}_{T-T}&=\sum g_{n_1n_1}g_{n_2n_2}g_{p_1p_1}g_{p_2p_2}\mathcal{H}_{D^*}(m_1)\mathcal{H}_{D^*}^*(m_2)\mathcal{M}_{T}(m_1,n_1,p_1)\mathcal{M}_{T}^*(m_2,n_2,p_2)\nl
        &\times  m_{\tau}^2
        \epsilon^{\mu}_{T}(n_1)
        \epsilon^{*\alpha}_{T}(n_2)
        \epsilon^{\nu}_{T}(p_1)
        \epsilon^{*\beta}_{T}(p_2)
        \epsilon^{*\xi}_{\rho}(k_1)
        \epsilon^{\zeta}_{\rho}(k_2)
        \nl&\times\Tr[\slashed{p}_{\nu_\tau}\gamma_{\xi}\sigma_{\mu\nu}(1-\gamma_5)\slashed{p}_{\bar\nu_\tau}\sigma_{\alpha\beta}\gamma_{\zeta}(1-\gamma_5)],
       \\[2ex]
        \mathcal{M}_{SP-VA}&=\sum g_{n_2n_2}\mathcal{H}_{D^*}(m_1)\mathcal{H}_{D^*}^*(m_2)\mathcal{M}_{SP}(m_1)\mathcal{M}_{VA}^*(m_2,n_2)\nl
        &\times  m_{\tau}
        \epsilon^{*\nu}_{VA}(n_2)
        \epsilon^{*\xi}_{\rho}(k_1)
        \epsilon^{\beta}_{\rho}(k_2)
        \Tr[\slashed{p}_{\nu_\tau}\gamma_{\xi}(1-\gamma_5)\slashed{p}_{\bar\nu_\tau}\gamma_{\nu}
        \slashed{p}_\tau \gamma_\beta (1-\gamma_5)],
   \\[2ex]
        \mathcal{M}_{SP-T}&=\sum
        g_{n_2n_2}g_{p_2p_2}
        \mathcal{H}_{D^*}(m_1)\mathcal{H}_{D^*}^*(m_2)
        \mathcal{M}_{SP}(m_1,n_1)
        \mathcal{M}_{T}^*(m_2,n_2,p_2)\nl
        &\times  m_{\tau}^2
        \epsilon^{*\alpha}_{T}(n_2)
        \epsilon^{*\beta}_{T}(k_2)
        \epsilon^{*\xi}_{\rho}(k_1)
        \epsilon^{\zeta}_{\rho}(k_2)
        \Tr[\slashed{p}_{\nu_\tau}
        \gamma_{\xi}
        (1-\gamma_5)
        \slashed{p}_{\bar\nu_\tau}
        \sigma_{\alpha\beta}
        \gamma_{\zeta}
        (1-\gamma_5)],
    \\[2ex]
        \mathcal{M}_{VA-SP}&=\sum
        g_{n_1n_1}
        \mathcal{H}_{D^*}(m_1)\mathcal{H}_{D^*}^*(m_2)
        \mathcal{M}_{VA}(m_1,n_1)
        \mathcal{M}_{SP}^*(m_2)\nl
        &\times  m_{\tau}
        \epsilon^{\mu}_{VA}(n_1)
        \epsilon^{*\alpha}_{\rho}(k_1)
        \epsilon^{\zeta}_{\rho}(k_2)
        \Tr[\slashed{p}_{\nu_\tau}
        \gamma_{\alpha}
        \slashed{p}_\tau
        \gamma_{\mu}
        (1-\gamma_5)
        \slashed{p}_{\bar\nu_\tau}
        \gamma_{\zeta}
        (1-\gamma_5)],
    \\[2ex]
        \mathcal{M}_{VA-T}&=\sum
        g_{n_1n_1}
        g_{n_2n_2}
        g_{p_2p_2}
        \mathcal{H}_{D^*}(m_1)\mathcal{H}_{D^*}^*(m_2)
        \mathcal{M}_{VA}(m_1,n_1)
        \mathcal{M}_{T}^*(m_2,n_2,p_2)\nl
        &\times  m_{\tau}
        \epsilon^{\mu}_{VA}(n_1)
        \epsilon^{*\alpha}_{T}(n_2)
        \epsilon^{*\beta}_{T}(p_2)
        \epsilon^{*\nu}_{\rho}(k_1)
        \epsilon^{\zeta}_{\rho}(k_2)
        \nl&\times\Tr[\slashed{p}_{\nu_\tau}
        \gamma_{\nu}
        \slashed{p}_\tau
        \gamma_{\mu}
        (1-\gamma_5)
        \slashed{p}_{\bar\nu_\tau}
        \sigma_{\alpha\beta}
        \gamma_{\zeta}
        (1-\gamma_5)],
   \\[2ex]
        \mathcal{M}_{T-SP}&=\sum
        g_{n_1n_1}
        g_{p_1p_1}
        \mathcal{H}_{D^*}(m_1)\mathcal{H}_{D^*}^*(m_2)
        \mathcal{M}_{T}(m_1,n_1,p_1)
        \mathcal{M}_{SP}^*(m_2)\nl
        &\times  m_{\tau}^2
        \epsilon^{\mu}_{T}(n_1)
        \epsilon^{\nu}_{T}(p_1)
        \epsilon^{*\xi}_{\rho}(k_1)
        \epsilon^{\zeta}_{\rho}(k_2)
        \Tr[\slashed{p}_{\nu_\tau}
        \gamma_{\xi}
        \sigma_{\mu\nu}
        (1-\gamma_5)
        \slashed{p}_{\bar\nu_\tau}
        \gamma_{\zeta}
        (1-\gamma_5)],
\\[2ex]
        \mathcal{M}_{T-VA}&=\sum
        g_{n_1n_1}
        g_{p_1p_1}
        g_{n_2n_2}
        \mathcal{H}_{D^*}(m_1)\mathcal{H}_{D^*}^*(m_2)
        \mathcal{M}_{T}(m_1,n_1,p_1)
        \mathcal{M}_{VA}^*(m_2,n_2)\nl
        &\times  m_{\tau}
        \epsilon^{\mu}_{T}(n_1)
       \epsilon^{*\alpha}_{VA}(n_2)
       \epsilon^{\nu}_{T}(p_1)
       \epsilon^{*\xi}_{\rho}(k_1)
       \epsilon^{\beta}_{\rho}(k_2)
        \Tr[\slashed{p}_{\nu_\tau}
        \gamma_{\xi}
        \sigma_{\mu\nu}
        (1-\gamma_5)
        \slashed{p}_{\bar\nu_\tau}
        \gamma_{\alpha}
        \slashed{p}_{\tau}
        \gamma_{\beta}
        (1-\gamma_5)].
\end{align}
The summation is taken over all helicity components, where $m_1$, $m_2$, $k_1$ and $k_2$ take the values $0$ and $\pm1$ while $n_1$, $n_2$, $p_1$ and $p_2$ take the values $0$, $\pm1$, and $t$.

 For the purpose of numerical illustration, we use the expressions presented in Ref.~\cite{1309.0301} for the form factors $V(q^2)$, $A_i(q^2)$ and $T_i(q^2)$, based on the heavy-quark effective theory (HQET). In fact, many relativistic calculations that do not rely on the heavy-quark approximations have been reported, in  quark models with the inclusion of relativistic effect \cite{Ivanov:2015tru,Cheng:2003sm,Faustov:2022ybm,Zhang:2020dla,Kang:2018jzg}. Note also that in Ref.~\cite{Faustov:2022ybm} the form factors can be calculated in the whole kinematic region without any extrapolation.
 The results of the relativistic quark model give $R_{D^*}=0.231$ or 0.246, deviating from the experimental value by more than $3\sigma$ \cite{Faustov:2022ybm} and see also Table IX therein.

\subsection{The full five-dimensional angular distribution}
After restoring the omitted factors, the modulus squared of the total amplitude is
\begin{equation}
\label{eq:MAbs2}
|\mathcal{M}|^2=\frac{G_{\text{F}}^2|V_{cb}|^2g_{D*D\pi}^2}{(p_{D^*}^2-m_{D^*}^2)^2+m_{D^*}^2\Gamma_{D^*}^2}\frac{G_{\text{F}}^2|V_{ud}|^2f_\rho^2 m_\rho^2}{(p_{\tau}^2-m_{\tau}^2)^2+m_{\tau}^2\Gamma_{\tau}^2}\sum_{k=\pm,0}\sum_\text{spins}|\mathcal{M}(k)|^2.
\end{equation}
The effective coupling constant of $D^*\rightarrow D\pi$, $g_{D^*D\pi}$, and $f_\rho$ can be obtained by their decay branching fractions:
\begin{align}
\mathcal{B}(D^*\rightarrow D\pi)&=\frac{g_{D^*D\pi}^2}{6\pi\: m_{D^*}^2\Gamma_{D^*}}|{\bf p}_{D}|^3,\\
\mathcal{B}(\tau^-\rightarrow\rho\nu_\tau)&=\frac{G_{\text{F}}^2|V_{ud}|^2f_\rho^2m_\tau^3}{16\pi\:\Gamma_\tau}
  \left(1-\frac{m_\rho^2}{m_\tau^2}\right)^2\left(1+\frac{2m_\rho^2}{m_\tau^2}\right).
\end{align}

Substituting them into Eq.~\eqref{eq:MAbs2}, one obtains
\begin{align}
|\mathcal{M}|^2&=\frac{96\pi^2\: G_{\text{F}}^2|V_{cb}|^2m_\rho^2}{|{\bf p}_D|^3(m_\tau^2-m_\rho^2)^2(m_\tau^2+2m_\rho^2)}
        \frac{\mathcal{B}(D^*\rightarrow D\pi)\Gamma_{D^*}m_{D^*}^2}{(p_{D^*}^2-m_{D^*}^2)^2+m_{D^*}^2\Gamma_{D^*}^2}
        \nl
        &\times\frac{\mathcal{B}(\tau^-\rightarrow\rho\nu_\tau)\Gamma_\tau m_\tau^3}{(p_{\tau}^2-m_{\tau}^2)^2+m_{\tau}^2\Gamma_{\tau}^2}\sum_{k=\pm,0}\sum_\text{spins}|\mathcal{M}(k)|^2.
\end{align}
 We observe that the dependence of $|\mathcal{M}|^2$ on ${\bf p}_{D^*}$ and ${\bf p}_{\tau}$ is only reflected in the propagator. With the narrow-width approximation, one has
\begin{equation}
    \int_{-\infty}^{+\infty} \frac{dp^2}{\pi}\frac{\mathcal{B}\:m\:\Gamma}{(p^2-m^2)^2+m^2\Gamma^2}\rightarrow\mathcal{B}.
\end{equation}
By integrating over the three variables ${\bf p}^2_{D^*}$, ${\bf p}^2_{\tau}$ and $\chi_{\tau\rho}$, the full five-body differential decay rate is given by
\begin{align}
\label{eq:d5Gamma_v1}
        \frac{d^5\Gamma}{d\cos{\theta^*dE_\rho d\cos{\theta_\rho d\chi_\rho d q^2}}}=\frac{3G_{\text{F}}^2|V_{cb}|^2|{\bf p}_{D^*}|\:m_\tau^4 (q^2)^{3/2}}{2^{11}\pi^{4}\:m_B^2(m_\tau^2-m_{\rho}^2)^2(m_\tau^2+2m_\rho^2)}\mathcal{B}(D^*\rightarrow D\pi)\nl
        \times \mathcal{B}(\tau^-\rightarrow\rho\nu_\tau)\sum_{i,j}\Big(\mathcal{C}_i^{S}|\mathcal{A}_i|^2+\mathcal{C}_{ij}^R\operatorname{Re}[\mathcal{A}_i\mathcal{A}_j^*]+\mathcal{C}_{ij}^I\operatorname{Im}[\mathcal{A}_i\mathcal{A}_j^*]\Big),
\end{align}
where $i,j=SP,\,0,\,t,\,\parallel,\,\perp,\, (0,T),\,(\parallel,T),\,(\perp,T)$, expressions of $\mathcal{A}_i$ have been given in Eq.~\eqref{eq:Ai}, $\mathcal{C}_i^S$, $\mathcal{C}_{ij}^R$, and $\mathcal{C}_{ij}^I$ are the coefficients, as functions of the five physical quantities ($q^2$, $E_\rho$, $\theta^*$, $\theta_{\rho}$, $\chi_\rho$). Their explicit expressions are all provided in Appendix \ref{amplitudecoefficient}.

Equation \eqref{eq:d5Gamma_v1} can be recast in to a clear form with angular functions and their coefficients:
\begin{align}
\label{angulardistribution}
   \frac{d^5\Gamma}{d\cos{\theta^*dE_\rho d\cos{\theta_\rho d\chi_\rho d q^2}}}=\frac{3G_{\text{F}}^2|V_{cb}|^2|{\bf p}_{D^*}|\:m_\tau^4 (q^2)^{3/2}}{2^{11}\pi^{4}\:m_B^2(m_\tau^2-m_{\rho}^2)^2(m_\tau^2+2m_\rho^2)}\mathcal{B}(D^*\rightarrow D\pi)\mathcal{B}(\tau^-\rightarrow\rho\nu_\tau)\nl
   \times\left(\sum_{i=1}^{9}X_i^R(q^2,E_\rho)\Omega_i^R(\theta^*,\theta_\rho,\chi_\rho)+\sum_{i=1}^{3}X_i^I(q^2,E_\rho)\Omega_i^I(\theta^*,\theta_\rho,\chi_\rho)\right).
\end{align}
The expressions of $X_i^R(q^2,E_\rho)$ and $\Omega_i^R(\theta^*,\theta_\rho,\chi_\rho)$ in Eq.~\eqref{angulardistribution} are provided in Table~\ref{R}, while the expressions of $X_i^I(q^2,E_\rho)$ and $\Omega_i^I(\theta^*,\theta_\rho,\chi_\rho)$ are given in Table~\ref{I}.

\begin{table}[t]
\centering
\caption{The structure of $X_i^R(q^2,E_\rho)$ and $\Omega_i^R(\theta^*,\theta_\rho,\chi_\rho)$.}
\label{R}
\begin{tabular}{|c|r|}
   \hline
    Coefficient $X_i^R(q^2,E_\rho)$ & Angular Function $\Omega_i^R(\theta^*,\theta_\rho,\chi_\rho)$\\
   \hline $\begin{aligned}\mathcal{S}_{SP}|\mathcal{A}_{SP}|^2+\mathcal{S}_{t}|\mathcal{A}_{t}|^2+\mathcal{S}_{0,1}|\mathcal{A}_0|^2
   +\mathcal{S}_{0T,1}|\mathcal{A}_{0T}|^2\\
   +\mathcal{R}_{SPt}\operatorname{Re}[\mathcal{A}_{SP}\mathcal{A}_t^*]+\mathcal{R}_{00T,1}\operatorname{Re}[\mathcal{A}_0\mathcal{A}_{0T}^*]\end{aligned}$
   &
   $\cos^2\theta^*$ \\
   \hline

    $\begin{aligned}\mathcal{R}_{t0}\operatorname{Re}[\mathcal{A}_{t}\mathcal{A}_{0}^*]+\mathcal{R}_{SP0}\operatorname{Re}[\mathcal{A}_{SP}\mathcal{A}_{0}^*]\\+\mathcal{R}_{SP0T}\operatorname{Re}[\mathcal{A}_{SP}\mathcal{A}_{0T}^*]+\mathcal{R}_{t0T}\operatorname{Re}[\mathcal{A}_{t}\mathcal{A}_{0T}^*]\end{aligned}$ & $\cos{\theta_\rho}\cos^2{\theta^*}$\\
     \hline

    $\begin{aligned}\mathcal{S}_{0,2}|\mathcal{A}_{0}|^2+\mathcal{S}_{0T,2}|\mathcal{A}_{0T}|^2+\mathcal{R}_{00T,2}\operatorname{Re}[\mathcal{A}_{0}\mathcal{A}_{0T}^*]\end{aligned}$ & $\cos{2\theta_\rho}\cos^2{\theta^*}$\\
    \hline

     $\begin{aligned}\mathcal{S}_{\parallel,1}|\mathcal{A}_{\parallel}|^2+\mathcal{S}_{\perp,1}|\mathcal{S}_{\perp}|^2+\mathcal{S}_{\parallel T,1}|\mathcal{A}_{\parallel T}|^2+\mathcal{S}_{\perp T,1}|\mathcal{A}_{\perp T}|^2\\
     +\mathcal{R}_{\parallel\parallel T,1 }\operatorname{Re}[\mathcal{A}_{\parallel}\mathcal{A}_{\parallel T}^*]+\mathcal{R}_{\perp\perp T,1 }\operatorname{Re}[\mathcal{A}_{\perp}\mathcal{A}_{\perp T}^*]\end{aligned}$ & $
     \sin^2\theta^*$\\
     \hline

     $\begin{aligned}\mathcal{R}_{\parallel\perp}\operatorname{Re}[\mathcal{A}_{\parallel}\mathcal{A}_{\perp}^*]+\mathcal{R}_{\parallel T\perp T}\operatorname{Re}[\mathcal{A}_{\parallel T}\mathcal{A}_{\perp T}^*]\\+\mathcal{R}_{\parallel\perp T}\operatorname{Re}[\mathcal{A}_{\parallel}\mathcal{A}_{\perp T}^*]+\mathcal{R}_{\perp\parallel T}\operatorname{Re}[\mathcal{A}_{\perp}\mathcal{A}_{\parallel T}^*]\end{aligned}$& $\cos{\theta_\rho}\sin^2{\theta^*}$\\
     \hline

      $\begin{aligned}\mathcal{S}_{\parallel,2}|\mathcal{A}_{\parallel}|^2+\mathcal{S}_{\perp,2}|\mathcal{A}_{\perp}|^2+\mathcal{S}_{\parallel T,2}|\mathcal{A}_{\parallel T}|^2+\mathcal{S}_{\perp T,2}|\mathcal{A}_{\perp T}|^2\\
      +\mathcal{R}_{\parallel\parallel T,2}\operatorname{Re}[\mathcal{A}_\parallel\mathcal{A}_{\parallel T}^*]+\mathcal{R}_{\perp\perp T,2}\operatorname{Re}[\mathcal{A}_\perp\mathcal{A}_{\perp T}^*]\end{aligned}$& $\cos{2\theta_\rho}\sin^2{\theta^*}$\\
     \hline

     $\begin{aligned}
         2(\mathcal{S}_{\perp,2}|\mathcal{A}_\perp|^2-\mathcal{S}_{\parallel,2}|\mathcal{A}_\parallel|^2+\mathcal{S}_{\perp T,2}|\mathcal{A}_{\perp T}|^2-\mathcal{S}_{\parallel T,2}|\mathcal{A}_{\parallel T}|^2)\\
         +2(\mathcal{R}_{\perp\perp T,2}\operatorname{Re}[\mathcal{A}_{\perp}\mathcal{A}_{\perp T}^*]-\mathcal{R}_{\parallel\parallel T,2}\operatorname{Re}[\mathcal{A}_{\parallel}\mathcal{A}_{\parallel T}^*])
     \end{aligned}$ & $\cos2\chi_\rho\sin^2\theta_\rho\sin^2{\theta^*}$\\
     \hline

     $\begin{aligned} \mathcal{R}_{t\parallel}\operatorname{Re}[\mathcal{A}_t\mathcal{A}_{\parallel}^*]
     +\mathcal{R}_{0\perp}\operatorname{Re}[\mathcal{A}_0\mathcal{A}_{\perp}^*]\\
     +\mathcal{R}_{0T\perp T}\operatorname{Re}[\mathcal{A}_{0T}\mathcal{A}_{\perp T}^*]+\mathcal{R}_{SP\parallel}\operatorname{Re}[\mathcal{A}_{SP}\mathcal{A}_{\parallel}^*]\\
     +\mathcal{R}_{SP\parallel T}\operatorname{Re}[\mathcal{A}_{SP}\mathcal{A}_{\parallel T}^*]+\mathcal{R}_{t\parallel T}\operatorname{Re}[\mathcal{A}_t\mathcal{A}_{\parallel T}^*]\\
     +\mathcal{R}_{0\perp T}\operatorname{Re}[\mathcal{A}_0\mathcal{A}_{\perp T}^*]+\mathcal{R}_{\perp 0T}\operatorname{Re}[\mathcal{A}_{\perp}\mathcal{A}_{0T}^*]\end{aligned}$&$\cos\chi_\rho\sin\theta_\rho\sin2\theta^*$\\

     \hline
     $\begin{aligned}
     \mathcal{R}_{0\parallel}\operatorname{Re}[\mathcal{A}_0\mathcal{A}_{\parallel}^*]+\mathcal{R}_{0T\parallel T}\operatorname{Re}[\mathcal{A}_{0T}\mathcal{A}_{\parallel T}^*]\\
     +\mathcal{R}_{0\parallel T}\operatorname{Re}[\mathcal{A}_{0}\mathcal{A}_{\parallel T}^*]+\mathcal{R}_{\parallel0T}\operatorname{Re}[\mathcal{A}_{\parallel}\mathcal{A}_{0T}^*]
     \end{aligned}$
     &
     $\cos\chi_\rho\sin2\theta_\rho\sin2\theta^*$\\
     \hline
\end{tabular}
\end{table}

\begin{table}[t]
\caption{The structure of $X_i^I(q^2,E_\rho)$ and $\Omega_i^I(\theta^*,\theta_\rho,\chi_\rho)$.}
     \label{I}
\centering
\begin{tabular}{|c|r|}
   \hline
   Coefficient $X_i^I(q^2,E_\rho)$ & Angular Function $\Omega_i^I(\theta^*,\theta_\rho,\chi_\rho)$\\
   \hline

   $\begin{aligned}
   \mathcal{I}_{t\perp}\operatorname{Im}[\mathcal{A}_{t}\mathcal{A}_{\perp}]+\mathcal{I}_{SP\perp}\operatorname{Im}[\mathcal{A}_{SP}\mathcal{A}_{\perp}]\\
   +\mathcal{I}_{SP\perp T}\operatorname{Im}[\mathcal{A}_{SP}\mathcal{A}_{\perp T}]+\mathcal{I}_{t\perp T}\operatorname{Im}[\mathcal{A}_{t}\mathcal{A}_{\perp T}]\\
   +\mathcal{I}_{0\parallel T}\operatorname{Im}[\mathcal{A}_{0}\mathcal{A}_{\parallel T}]+\mathcal{I}_{\parallel 0T}\operatorname{Im}[\mathcal{A}_{\parallel}\mathcal{A}_{0 T}]
   \end{aligned}$&$\sin\chi_\rho\sin\theta_\rho\sin2\theta^*$\\
   \hline

   $\begin{aligned}
       \mathcal{I}_{0\perp}\operatorname{Im}[\mathcal{A}_{0}\mathcal{A}_{\perp}]+\mathcal{I}_{0\perp T}\operatorname{Im}[\mathcal{A}_{0}\mathcal{A}_{\perp T}]\\
       +\mathcal{I}_{\perp0T}\operatorname{Im}[\mathcal{A}_{\perp}\mathcal{A}_{0T}]
   \end{aligned}$&$\sin\chi_\rho\sin2\theta_\rho\sin2\theta^*$\\
   \hline

   $\begin{aligned}
       \mathcal{I}_{\parallel\perp}\operatorname{Im}[\mathcal{A}_{\parallel}\mathcal{A}_{\perp}]+\mathcal{I}_{\parallel\perp T}\operatorname{Im}[\mathcal{A}_{\parallel}\mathcal{A}_{\perp T}]\\
       +\mathcal{I}_{\perp\parallel T}\operatorname{Im}[\mathcal{A}_{\perp}\mathcal{A}_{\parallel T}]
   \end{aligned}$&$\sin2\chi_\rho\sin^2\theta_\rho\sin^2\theta^*$\\
   \hline
\end{tabular}
\end{table}

Apart from the $\tau^-\rightarrow \rho^-\nu_\tau$ decay channel, our such angular analysis can be extended to the $\tau^-\rightarrow a_1^-\nu_\tau$ channel, which accounts for most of the $\tau\rightarrow 3\pi\nu_\tau$ decay branching fraction. Nevertheless, needless to say, the helicity angles related to the $\rho$ should be replaced by the ones describing the $a_1$ decay orientation.

Table~\ref{R} contains terms that are already present in the SM, for e.g., the helicity amplitudes $\mathcal{A}_t$, $\mathcal{A}_0$, $\mathcal{A}_\parallel$ and $\mathcal{A}_\perp$. New physics modifies these coupling coefficients in two distinct ways. First, the amplitudes depend on the combinations $1+g_{L}-g_{R}$ and $1+g_{L}+g_{R}$. By performing a detailed angular analysis, any fitted values of $g_L$ and/or $g_R$ away from 0 signifies the existence of NP contributions. Second, the NP amplitudes can interfere with the SM amplitudes. As an example, the scalar amplitude $\mathcal{A}_{SP}$ generated by $g_{P}$ interferes with the longitudinal amplitudes $\mathcal{A}_0$, producing terms such as $\operatorname{Re}[\mathcal{A}_{SP}\mathcal{A}_{0}^*]$. Similarly, the tensor amplitudes $\mathcal{A}_{0,T}$, $\mathcal{A}_{\parallel,T}$, and $\mathcal{A}_{\perp,T}$ generated by $g_T$ also interfere with the SM ones. The nonzero contributions of these terms as well indicate the existence of NP.

All entries of angular terms in Table ~\ref{I} depend on $\sin\chi_\rho$ or $\sin2\chi_\rho$,
which can be expressed as
\begin{align}
\label{eq:sinchirho}
\sin\chi_\rho&=\frac{[({\bf p}_\pi\times {\bf p}_D)\times ({\bf p}_{D^*}\times{\bf p}_\rho)]\cdot {\bf p}_{D^*}}{|{\bf p}_\pi\times {\bf p}_D| |{\bf p}_{D^*}\times{\bf p}_\rho| |{\bf p}_{D^*}|}\nl
&\propto ({\bf p}_\pi\times {\bf p}_D)\cdot {\bf p}_\rho.
\end{align}
Or in other words, $\sin\chi_\rho$ indicates a triple-product asymmetry, which changes sign under time reversal.
The coefficients involve the imaginary part of the product of two helicity amplitudes, $\operatorname{Im}[\mathcal{A}_i\mathcal{A}_j^*]$. We write a general amplitude as $\mathcal{A}=|\mathcal{A}|e^{i\left(\phi+\delta\right)}$, where $\phi$ is the weak (CP?odd) phase and $\delta$ strong (CP-even) phase. Then
\begin{equation}
\text{Im}\,[\mathcal{A}_i\mathcal{A}_j^*]=|\mathcal{A}_i||\mathcal{A}_j|\sin\left[\left({\phi_i}-\phi_j\right)+\left({\delta_i}-\delta_j\right)\right].
\end{equation}
However, one should be cautious that a nonzero $\text{Im}\,[\mathcal{A}_i\mathcal{A}_j^*]$  can arise either from a weak-phase difference $\left({\phi_i}-\phi_j\right)$ or a strong-phase difference $\left(\delta_i-\delta_j\right)$ alone and thus does not necessarily imply a true CP-violating signal \cite{Bigibook,Valencia,Datta2003,Duraisamy:2013pia}.  A genuine CP-violating signal requires a comparison between the $\bar{B}$ decay and its CP-conjugate $B$ decay \footnote{The sensitivities of these triple-product asymmetry for the strange and charm quark sector at BESIII or super tau-charm factory are  discussed in Refs.~\cite{Bigi:2017eni,Shi:2019vus,Kang:2010td,Kang:2009iy}.}. The latter contains the term $-\sin\chi_\rho$ (the minus sign is due to the triple product form in Eq.~\eqref{eq:sinchirho})  and $\text{Im}\,[\bar{\mathcal{A}}_i\bar{\mathcal{A}}_j^*]$. Putting the minus in front of $\text{Im}\,[\bar{\mathcal{A}}_i\bar{\mathcal{A}}_j^*]$, one has
\begin{equation}
-\text{Im}\,[\bar{\mathcal{A}}_i\bar{\mathcal{A}}_j^*]=|\mathcal{A}_i||\mathcal{A}_j|\sin\left[\left({\phi_i}-\phi_j\right)-\left({\delta_i}-\delta_j\right)\right].
\end{equation}
The true CP-violating term is thus constructed by adding the TP asymmetry in the process and its CP-conjugate process, such that the T-violating term is proportional to $\sin(\phi_i-\phi_j)\cos(\delta_i-\delta_j)$. In a practical angular analysis, one just identifies the coefficient of $\sin\chi_\rho$ term in an untagged data sample, which is a true CP violating observable. It is important to note that  CP violation is absent in the channel $\bar B\to D^*\tau\nu$ for the SM, and thus an experimental observation of the entries in Table ~\ref{I} would definitely be a smoking-gun signal of NP.

\section{Integrated Observables}\label{Integrated Observables}

The decay $\bar B\rightarrow D^*(\rightarrow D\pi)\tau^-(\rightarrow \rho^-\nu_\tau)\bar\nu_\tau$ depends on five physical quantities, namely $\theta^*$, $\theta_\rho$, $\chi_\rho$, $E_\rho$ and $q^2$.
In principle, it is always desirable to determine the dynamics by performing a full angular analysis. However, in reality, such an analysis is limited by statistical samples.
In such cases, one can define useful observables to study the NP effect by integrating the differential decay rate over one or more kinematic variables.

We can categorize the kinetic parameters into two groups. One group consists of $\theta_\rho$, $\chi_\rho$, and $E_\rho$, associated with the leptonic side of the decay chain. The second category consists of $q^2$ and $\theta^*$ that are related to the  hadronic side.

By integrating over $E_\rho$, $\cos\theta_\rho$ and $\chi_\rho$, we obtain
\begin{align}
    \frac{d^2 \Gamma}{ d\cos\theta^*   d q^2}&=\int_{E_{\text{min}}}^{E_{\text{max}}}\int_{-1}^{1}\int_0^{2\pi}\frac{d^5 \Gamma}{ d\cos\theta^* dE_\rho d \cos \theta_\rho d\chi_\rho d q^2} d\chi_\rho d\cos\theta_\rho d E_\rho,
\end{align}
where the range of $E_\rho$ is given by
\begin{equation}
    \frac{m_\tau^4+m^2_\rho q^2}{2m_\tau^2\sqrt{q^2}}\le E_\rho \le \frac{q^2+m_\rho^2}{2\sqrt{q^2}}.
\end{equation}
Then we have
\begin{align}
    \frac{d^2 \Gamma}{ d\cos\theta^*   d q^2}&=A(q^2)\cos^2\theta^*+B(q^2)\sin^2\theta^*\nonumber\\
    &=\frac{3}{4}\frac{d \Gamma}{d q^2}[2F_L^{D^*}\cos^2\theta^*+F_T^{D^*}\sin^2\theta^*],
\end{align}
with
\begin{align}
\label{eq:dGammadq2}
    \frac{d \Gamma}{d q^2}=\frac{2}{3}A(q^2)+\frac{4}{3}B(q^2),
\end{align}
and
\begin{align}
    A(q^2)&=\frac{ G_{\text{F}}^2 |V_{cb}|^2 |{\bf p}_{D^*}|q^2}{2^7\pi^3m_B^2}\left(\frac{m_\tau^2}{q^2}-1\right)^2\mathcal{B}(D^*\rightarrow D\pi)\:\mathcal{B}(\tau^-\rightarrow\rho\nu_\tau)
    \Bigg(
    \left(\frac{m_\tau^2}{q^2}+2\right)|\mathcal{A}_0|^2
    \nl&+16 \left(\frac{2 m_\tau^2}{q^2}+1\right)|\mathcal{A}_{0T}|^2
    +3\Big(\mathcal{A}_{SP}
    +\frac{m_\tau}{\sqrt{q^2}}\mathcal{A}_t\Big)^2
    -24\frac{m_\tau}{\sqrt{q^2}}\operatorname{Re}[\mathcal{A}_0\mathcal{A}_{0T}^*]
    \Bigg),\\
    B(q^2)&=\frac{G_{\text{F}}^2 |V_{cb}|^2 |{\bf p}_{D^*}|q^2}{2^8\pi^3 m_B^2 }\left(\frac{m_\tau^2}{q^2}-1\right)^2\mathcal{B}(D^*\rightarrow D\pi)\:\mathcal{B}(\tau^-\rightarrow\rho\nu_\tau)
   \Bigg(
    \left(\frac{m_\tau^2}{q^2}+2\right)\left(|\mathcal{A}_{\parallel}|^2+|\mathcal{A}_{\perp}|^2\right)
    \nl
    &+16\left(2\frac{m_\tau^2}{q^2}+1\right)\left(|\mathcal{A}_{\parallel T}|^2+|\mathcal{A}_{\perp T}|^2\right)
    -24\frac{m_\tau}{\sqrt{q^2}}
    \left(\operatorname{Re}[\mathcal{A}_\parallel\mathcal{A}_{\parallel T}]+\operatorname{Re}[\mathcal{A}_\perp\mathcal{A}_{\perp T}]\right)
    \Bigg).
\end{align}
The expressions for $F_L^{D^*}$ and $F_T^{D^*}$ are written as
\begin{align}
    F_L^{D^*}=\frac{A(q^2)}{A(q^2)+2B(q^2)},\quad F_T^{D^*}=\frac{2B(q^2)}{A(q^2)+2B(q^2)},\quad F_L^{D^*}+F_T^{D^*}=1,
\end{align}
which correspond to the longitudinal and transverse polarization fractions of the $D^*$.
We comment that Eq.~\eqref{eq:dGammadq2} agrees with Eq.~(4.5) in Ref.~\cite{2005},
with replacing $\mathcal{B}(\tau\to\pi \nu_\tau)$ therein by $\mathcal{B}(\tau\to\rho \nu_\tau)$ in our case. One can compute the ratio $R_{D^*}$ by  the integration of Eq.~\eqref{eq:dGammadq2}.

The observable related to the leptonic side, following the $\tau\rightarrow\rho\nu_\tau$ decay, can be obtained by integration over $\theta^*$. As a result, we define
\begin{equation}
    \frac{d^3 \Gamma}{ dE_\rho d \cos \theta_\rho  d q^2}=\int_{0}^{2\pi}\int_{-1}^{1}\frac{d^5 \Gamma}{ d\cos\theta^* dE_\rho d \cos \theta_\rho d\chi_\rho d q^2}d\cos\theta^*d\chi_\rho,
\end{equation}
and obtain
\begin{align}
    \frac{d^3\Gamma}{dE_\rho d\cos\theta_\rho dq^2}&=a(E_\rho,q^2)+b(E_\rho,q^2)\cos\theta_\rho+c(E_\rho,q^2)\cos2\theta_\rho\nonumber\\
    &=\frac{3}{2}\frac{d^2\Gamma}{dE_\rho dq^2}\frac{a(E_\rho,q^2)+b(E_\rho,q^2)\cos\theta_\rho+c(E_\rho,q^2)\cos2\theta_\rho}{3 a(E_\rho,q^2)-c(E_\rho,q^2)},
\end{align}
with
\begin{align}
    \frac{d^2\Gamma}{dE_\rho dq^2}&=\int_{-1}^{1}[a(E_\rho,q^2)+b(E_\rho,q^2)\cos\theta_\rho+c(E_\rho,q^2)\cos2\theta_\rho]d\cos\theta_\rho\nl
    &=\frac{2}{3} \big(3a(E_\rho,q^2)- c(E_\rho,q^2)\big),
    \\[2ex]
    a(E_\rho,q^2)&=\frac{G_{\text{F}}^2|V_{cb}|^2|{\bf p}_{D^*}|m_\tau^4(q^2)^{3/2}}{2^9 \pi^3m_B^2(m_\tau^2-m_\rho^2)^2(m_\tau^2+2m_\rho^2)}\mathcal{B}(D^*\rightarrow D\pi)\mathcal{B}(\tau^-\rightarrow\rho\nu_\tau)
    \nonumber\\
    &\Big\{
    \mathcal{S}_{SP}|\mathcal{A}_{SP}|^2
    +\mathcal{S}_{t}|\mathcal{A}_{t}|^2
    +2\mathcal{S}_{\parallel,1}|\mathcal{A}_{\parallel}|^2
    +2\mathcal{S}_{\perp,1}|\mathcal{A}_{\perp}|^2
    +2\mathcal{S}_{\parallel T,1}|\mathcal{A}_{\parallel T}|^2
    +2\mathcal{S}_{\perp T,1}|\mathcal{A}_{\perp T}|^2\nonumber\\
   &+\mathcal{S}_{0,1}|\mathcal{A}_{0}|^2
    +\mathcal{S}_{0T,1}|\mathcal{A}_{0T}|^2
    +\mathcal{R}_{SPt}\operatorname{Re}[\mathcal{A}_{SP}\mathcal{A}_t^*]
    +\mathcal{R}_{00T,1}\operatorname{Re}[\mathcal{A}_{0}\mathcal{A}_{0T}^*]\nonumber\\
    &+2\mathcal{R}_{\parallel\parallel T,1}\operatorname{Re}[\mathcal{A}_{\parallel}\mathcal{A}_{\parallel T}^*]
    +2\mathcal{R}_{\perp\perp T,1}\operatorname{Re}[\mathcal{A}_{\perp}\mathcal{A}_{\perp T}^*]
    \Big\},
    \\[2ex]
    b(E_\rho,q^2)&=\frac{G_{\text{F}}^2|V_{cb}|^2|{\bf p}_{D^*}|m_\tau^4(q^2)^{3/2}}{2^9 \pi^3m_B^2(m_\tau^2-m_\rho^2)^2(m_\tau^2+2m_\rho^2)}\mathcal{B}(D^*\rightarrow D\pi)\mathcal{B}(\tau^-\rightarrow\rho\nu_\tau)
    \Big\{
    \mathcal{R}_{SP0}\operatorname{Re}[\mathcal{A}_{SP}\mathcal{A}_{0}^*]\nonumber\\
    &+\mathcal{R}_{SP0T}\operatorname{Re}[\mathcal{A}_{SP}\mathcal{A}_{0T}^*]
    +\mathcal{R}_{t0}\operatorname{Re}[\mathcal{A}_{t}\mathcal{A}_{0}^*]
    +\mathcal{R}_{t0T}\operatorname{Re}[\mathcal{A}_{t}\mathcal{A}_{0T}^*]
    +2\mathcal{R}_{\parallel\perp}\operatorname{Re}[\mathcal{A}_{\parallel}\mathcal{A}_\perp^*]\nonumber\\
    &+2\mathcal{R}_{\parallel T\perp T}\operatorname{Re}[\mathcal{A}_{\parallel T}\mathcal{A}_{\perp T}^*]
    +2\mathcal{R}_{\parallel T\perp}\operatorname{Re}[\mathcal{A}_{\parallel T}\mathcal{A}_\perp^*]
    +2\mathcal{R}_{\parallel\perp T}\operatorname{Re}[\mathcal{A}_{\parallel}\mathcal{A}_{\perp T}^*]
    \Big\},
    \\[2ex]
    c(E_\rho,q^2)&=\frac{G_{\text{F}}^2|V_{cb}|^2|{\bf p}_{D^*}|m_\tau^4(q^2)^{3/2}}{2^9 \pi^3m_B^2(m_\tau^2-m_\rho^2)^2(m_\tau^2+2m_\rho^2)}\mathcal{B}(D^*\rightarrow D\pi)\mathcal{B}(\tau^-\rightarrow\rho\nu_\tau)
    \nonumber\\
    &\Big\{
    \mathcal{S}_{0,2}|\mathcal{A}_0|^2
    +\mathcal{S}_{0T,2}|\mathcal{A}_{0T}|^2
    +2\mathcal{S}_{\parallel,2}|\mathcal{A}_{\parallel}|^2
    +2\mathcal{S}_{\perp,2}|\mathcal{A}_{\perp}|^2
    +2\mathcal{S}_{\parallel T,2}|\mathcal{A}_{\parallel T}|^2
    +2\mathcal{S}_{\perp T,2}|\mathcal{A}_{\perp T}|^2\nonumber\\
    &+\mathcal{R}_{00T,2}\operatorname{Re}[\mathcal{A}_0\mathcal{A}_{0 T}^*]
    +2\mathcal{R}_{\parallel\parallel T,2}\operatorname{Re}[\mathcal{A}_{\parallel}\mathcal{A}_{\parallel T}^*]
    +2\mathcal{R}_{\perp\perp T,2}\operatorname{Re}[\mathcal{A}_{\perp}\mathcal{A}_{\perp T}^*]
    \Big\}.
\end{align}

From the $\theta_\rho$ distribution, one can define the forward-backward asymmetry as follows:
\begin{align}    \eta_1(E_\rho,q^2)=&\frac{\int_{-1}^0\frac{d^3\Gamma}{dE_\rho d\cos\theta_\rho dq^2}d\cos\theta_\rho-\int_{0}^{1}\frac{d^3\Gamma}{dE_\rho d\cos\theta_\rho dq^2}d\cos\theta_\rho}{\frac{d^2\Gamma}{dE_\rho dq^2}}\nonumber\\
    =&-\frac{3}{2}\frac{b(E_\rho,q^2)}{3a(E_\rho,q^2)-c(E_\rho,q^2)},
\end{align}
which is a parity-violating observable is nonzero even in the SM.   .

Similarly, one can study the azimuthal distribution in $\chi_\rho$. To that end, we have
\begin{align}
    \frac{d^3\Gamma}{dE_\rho d\chi_\rho dq^2}&=\int_{-1}^{1}\int_{-1}^{1}\frac{d^5 \Gamma}{ d\cos\theta^* dE_\rho d \cos \theta_\rho d\chi_\rho d q^2}d\cos\theta^*d\cos\theta_\rho
    \nl  &=f(E_\rho,q^2)+g(E_\rho,q^2)\cos2\chi_\rho+h(E_\rho,q^2)\sin2\chi_\rho\nonumber\\
    &=\frac{1}{2\pi}\frac{d^2\Gamma}{dE_\rho dq^2}\frac{f(E_\rho,q^2)+g(E_\rho,q^2)\cos2\chi_\rho+h(E_\rho,q^2)\sin2\chi_\rho}{f(E_\rho,q^2)},
\end{align}
with
\begin{align}
    \frac{d^2\Gamma}{dE_\rho dq^2}&=2\pi f(E_\rho,q^2),\\
    f(E_\rho,q^2)&=\frac{1}{3}\frac{G_{\text{F}}^2|V_{cb}|^2|{\bf p}_{D^*}|m_\tau^4(q^2)^{3/2}}{2^{9} \pi^4 m_B^2(m_\tau^2-m_\rho^2)^2(m_\tau^2+2m_\rho^2)}\mathcal{B}(D^*\rightarrow D\pi)\mathcal{B}(\tau^-\rightarrow\rho\nu_\tau)\Big\{3\mathcal{S}_{SP}|\mathcal{A}_{SP}|^2
    \nonumber\\
    &+3\mathcal{S}_{t}|\mathcal{A}_{t}|^2
    +\left(3\mathcal{S}_{0,1}-\mathcal{S}_{0,2}\right)|\mathcal{A}_0|^2
    +\left(3\mathcal{S}_{0T,1}-\mathcal{S}_{0T,2}\right)|\mathcal{A}_{0T}|^2+
    \left(6\mathcal{S}_{\parallel,1}-2\mathcal{S}_{\parallel,2}\right)|\mathcal{A}_{\parallel}|^2
    \nonumber\\
    &+\left(6\mathcal{S}_{\perp,1}-2\mathcal{S}_{\perp,2}\right)|\mathcal{A}_{\perp}|^2
    +\left(6\mathcal{S}_{\parallel T,1}-2\mathcal{S}_{\parallel T,2}\right)|\mathcal{A}_{\parallel T}|^2
    +\left(6\mathcal{S}_{\perp T,1}-2\mathcal{S}_{\perp T,2}\right)|\mathcal{A}_{\perp T}|^2
    \nonumber\\
    &+\left(6\mathcal{R}_{\parallel\parallel T,1}-2\mathcal{R}_{\parallel\parallel T,2}\right)\operatorname{Re}[\mathcal{A}_\parallel\mathcal{A}_{\parallel T}^*]
    +\left(6\mathcal{R}_{\perp\perp T,1}-2\mathcal{R}_{\perp\perp T,2}\right)\operatorname{Re}[\mathcal{A}_\perp\mathcal{A}_{\perp T}^*]\nonumber\\
    &
    +3\mathcal{R}_{SPt}\operatorname{Re}[\mathcal{A}_{SP}\mathcal{A}_{t}^*]
    +\left(3\mathcal{R}_{00T,1}-\mathcal{R}_{00T,2}\right)\operatorname{Re}[\mathcal{A}_{0}\mathcal{A}_{0T}^*]
    \Big\},\\[2ex]
    g(E_\rho,q^2)&=\frac{1}{3}\frac{G_{\text{F}}^2|V_{cb}|^2|{\bf p}_{D^*}|m_\tau^4(q^2)^{3/2}}{2^{6} \pi^4 m_B^2(m_\tau^2-m_\rho^2)^2(m_\tau^2+2m_\rho^2)}\mathcal{B}(D^*\rightarrow D\pi)\mathcal{B}(\tau^-\rightarrow\rho\nu_\tau)
    \Big\{\mathcal{S}_{\perp,2}|\mathcal{A}_{\perp}|^2
    -\mathcal{S}_{\parallel,2}|\mathcal{A}_{\parallel}|^2\nonumber\\
    &
    +\mathcal{S}_{\perp T,2}|\mathcal{A}_{\perp T}|^2
    -\mathcal{S}_{\parallel T,2}|\mathcal{A}_{\parallel T}|^2
    +\mathcal{R}_{\perp\perp T,2}\operatorname{Re}[\mathcal{A}_{\perp}\mathcal{A}_{\perp T}^*]
    -\mathcal{R}_{\parallel\parallel T,2}\operatorname{Re}[\mathcal{A}_{\parallel}\mathcal{A}_{\parallel T}^*]\Big\},\\[2ex]
    h(E_\rho,q^2)&=\frac{1}{3}\frac{G_{\text{F}}^2|V_{cb}|^2|{\bf p}_{D^*}|m_\tau^4(q^2)^{3/2}}{2^{7} \pi^4 m_B^2(m_\tau^2-m_\rho^2)^2(m_\tau^2+2m_\rho^2)}\mathcal{B}(D^*\rightarrow D\pi)\mathcal{B}(\tau^-\rightarrow\rho\nu_\tau)
    \nonumber\\
    &\Big\{
    \mathcal{I}_{\parallel\perp}\operatorname{Im}[\mathcal{A}_{\parallel}\mathcal{A}_{\perp}^*]
    +\mathcal{I}_{\parallel\perp T}\operatorname{Im}[\mathcal{A}_{\parallel}\mathcal{A}_{\perp T}^*]
    +\mathcal{I}_{\perp\parallel T}\operatorname{Im}[\mathcal{A}_{\perp}\mathcal{A}_{\parallel T}^*]\Big\}.
\end{align}

 The asymmetric observable corresponding to $\chi_\rho$ is given by
\begin{align}
    \eta_2(E_\rho,q^2)&=\frac{\left(\int_0^{\pi/2}-\int_{\pi/2}^\pi+\int_{\pi}^{3\pi/2}-\int_{3\pi/2}^{2\pi}\right)\frac{d^3 \Gamma}{dE_\rho d\chi_\rho dq^2}d\chi_\rho}{\frac{d^2 \Gamma}{dE_\rho dq^2}}\nonumber\\
    &=\frac{2}{\pi}\frac{h(E_\rho,q^2)}{f(E_\rho,q^2)},
\end{align}
which corresponds to the distribution asymmetry by $\sin 2\chi_\rho>0$ and  $\sin 2\chi_\rho<0$. Note the $h$ term contains the imaginary part of the interference of helicity amplitudes, such as $\text{Im}\,[\mathcal{A}_\parallel\mathcal{A}_{\perp}^*]$, which is the time-reversal-violating triple-product asymmetry. As mentioned earlier, if one  measures $\eta_2 \neq 0$ in an untagged data sample then it is a clear signal of  CP violation  as well as NP.


\section{Extracting the $\tau$ polarization from $\rho$ energy spectra}\label{pp}

In Sec.~\ref{Integrated Observables}, we presented various integrated observables from the full five-dimensional angular distribution of the decay chain. In this section, we show that the spin structure of  the $\rho$ in the $\tau\rightarrow\rho\nu_\tau$ decay allows the decay chain to be decomposed into the longitudinal ($L$) and transverse ($T$) helicity sectors of the $\rho$, and the $\tau$  polarization $P_\tau^{D^*}$ can be extracted directly from the slope of the resulting helicity-separated $\rho$ energy spectra. We assume that new physics affects only the production-side $b\rightarrow c\tau\bar{\nu}_\tau$ transition, while the $\tau$ decay is governed by the left-handed $V-A$ interaction of the Standard Model.

\begin{figure}
    \centering
    \includegraphics[width=0.5\linewidth]{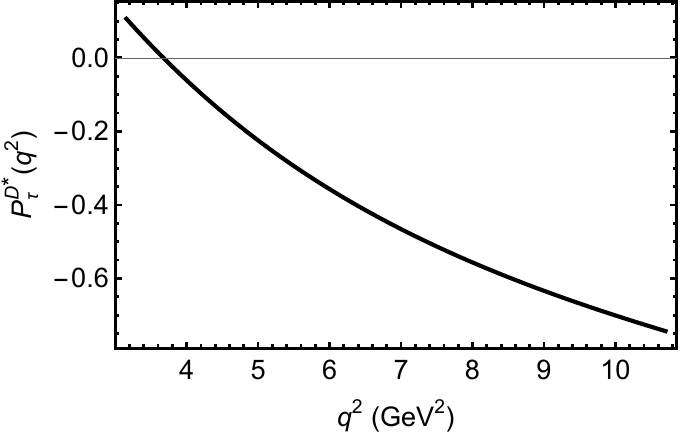}
    \caption{Plot for $P_\tau^{D^*}$ as a function of $q^2$ in the framework of Standard Model.}
    \label{Ptau}
\end{figure}

After integrating over the relevant angles, the differential decay rate of the total decay chain can be written as (splitting the amplitude and phase space, see e.g., in Ref.~\cite{Kang:2016jxw})
\begin{align}
\label{dGammadq2dErho}
    \frac{d^2\Gamma_h}{dq^2dE_\rho}=\frac{1}{\Gamma_\tau}\sum_{\lambda=\uparrow,\downarrow}\frac{d\Gamma(\bar{B}\rightarrow D^*(\to D\pi)\tau_\lambda\bar{\nu}_\tau)}{dq^2}\frac{d\Gamma(\tau_\lambda\rightarrow\rho_h\nu_\tau)}{dE_\rho},\quad h=L,T,
\end{align}
where $L$ and $T$ represent the longitudinal and transverse components of the helical degree of $\rho$, respectively. And $\Gamma_\tau$ is the total $\tau$ width, $\lambda=\uparrow, \downarrow$ is the helicity of $\tau$. The complete expressions of $\frac{d\Gamma(\bar{B}\rightarrow D^* (\to D\pi)\tau_\lambda\bar{\nu}_\tau)}{d q^2}$ are
\begin{align}
    \frac{d\Gamma(\bar{B}\rightarrow D^*(\to D\pi)\tau_{\uparrow}\bar{\nu}_\tau)}{dq^2}&=\frac{1}{3}\frac{G_F^2|V_{cb}|^2|{\bf p}_{D^*}| q^2}{2^6\pi^3m_B^2}\left(\frac{m_\tau^2}{q^2}-1\right)^2\mathcal{B}(D^*\rightarrow D\pi)
    \Bigg(
    3\Big(\mathcal{A}_{SP}+\frac{m_\tau}{\sqrt{q^2}}\mathcal{A}_{t}\Big)^2\nonumber\\
    &+16\left(|\mathcal{A}_{0T}|^2+|\mathcal{A}_{\parallel T}|^2+|\mathcal{A}_{\perp T}|^2\right)+\frac{m_\tau^2}{q^2}\left(|\mathcal{A}_0|^2+|\mathcal{A}_{\parallel}|^2+|\mathcal{A}_{\perp}|^2\right)\nonumber\\
    &-8\frac{m_\tau}{\sqrt{q^2}}\left(\operatorname{Re}[\mathcal{A}_{0}\mathcal{A}_{0T}^*]+\operatorname{Re}[\mathcal{A}_{\parallel}\mathcal{A}_{\parallel T}^*]+\operatorname{Re}[\mathcal{A}_{\perp}\mathcal{A}_{\perp T}^*]\right)
    \Bigg),
\end{align}
\begin{align}
   \frac{d\Gamma(\bar{B}\rightarrow D^*(\to D\pi)\tau_{\downarrow}\bar{\nu}_\tau)}{dq^2}&=\frac{1}{3}\frac{G_F^2|V_{cb}|^2|{\bf p}_{D^*}| q^2}{2^5\pi^3m_B^2}\left(\frac{m_\tau^2}{q^2}-1\right)^2\mathcal{B}(D^*\rightarrow D\pi)
    \Bigg(
    \left(|\mathcal{A}_0|^2+|\mathcal{A}_{\parallel}|^2+|\mathcal{A}_{\perp}|^2\right)\nonumber\\
    &+16\frac{m_\tau^2}{q^2}\left(|\mathcal{A}_{0T}|^2+|\mathcal{A}_{\parallel T}|^2+|\mathcal{A}_{\perp T}|^2\right)\nonumber\\
    &-8\frac{m_\tau}{\sqrt{q^2}}\left(\operatorname{Re}[\mathcal{A}_{0}\mathcal{A}_{0T}^*]+\operatorname{Re}[\mathcal{A}_{\parallel}\mathcal{A}_{\parallel T}^*]+\operatorname{Re}[\mathcal{A}_{\perp}\mathcal{A}_{\perp T}^*]\right)
    \Bigg).
\end{align}
It can be verified that
\begin{align}
    \frac{d\Gamma(\bar{B}\rightarrow D^*(\to D\pi)\tau_{\uparrow}\bar{\nu}_\tau)}{dq^2}+\frac{d\Gamma(\bar{B}\rightarrow D^*(\to D\pi)\tau_{\downarrow}\bar{\nu}_\tau)}{dq^2}=\frac{1}{\mathcal{B}(\tau\rightarrow\rho\nu_\tau)}\frac{d\Gamma}{dq^2},
\end{align}
with $\frac{d\Gamma}{dq^2}$ given by Eq.~\eqref{eq:dGammadq2} from the five-body decay chain.
In this case, the explicit expression for the $\tau$ polarization, $P_{\tau}^{D^*}(q^2)$, is
\begin{align}
    P_{\tau}^{D^*}(q^2)=\mathcal{B}(\tau\rightarrow\rho\nu_\tau)\frac{\frac{d\Gamma(\bar{B}\rightarrow D^*(\to D\pi)\tau_{\uparrow}\bar{\nu}_\tau)}{dq^2}-\frac{d\Gamma(\bar{B}\rightarrow D^*(\to D\pi)\tau_{\downarrow}\bar{\nu}_\tau)}{dq^2}}{\frac{d\Gamma}{dq^2}}.
\end{align}
We draw in Fig.~\ref{Ptau} for $P_\tau^{D^*}$ as a function of $q^2$ under the SM framework.
The value of $P_\tau^{D^*}$ decreases as $q^2$ increases, dropping from approximately $+0.10$ to $-0.74$.

Substituting the above expression of $P_\tau^{D^*}(q^2)$ into Eq.~\eqref{dGammadq2dErho}, we obtain
\begin{align}
\label{eq:dGhPtau}
    \frac{d^2\Gamma_h}{dq^2dE_\rho}&=\frac{1}{\Gamma_\tau \mathcal{B}(\tau\rightarrow\rho\nu_\tau)}\frac{d\Gamma}{dq^2}\left(\frac{1+P_{\tau}^{D^*}(q^2)}{2}
    \frac{d\Gamma\left(\tau_\uparrow\rightarrow\rho_h \nu_\tau\right)}{dE_\rho}+\frac{1-P_{\tau}^{D^*}(q^2)}{2}\frac{d\Gamma\left(\tau_\downarrow\rightarrow\rho_h \nu_\tau\right)}{dE_\rho}\right).
\end{align}
The above Eq.~\eqref{eq:dGhPtau} will be finally expressed as
\begin{align}
\label{YYEQ}
    \frac{d^2\Gamma_h}{dq^2dE_\rho}
    =
    \frac{d\Gamma}{dq^2}\left(\bar{Y}_h(q^2,E_\rho)+P_{\tau}^{D^*}(q^2)\Delta Y_h(q^2,E_\rho)\right).
\end{align}
Let us focus on the discussion of $P_\tau^{D^*}$ and avoid digressing from it. The specific expressions for $\frac{d\Gamma\left(\tau_\lambda\rightarrow\rho_h \nu_\tau\right)}{dE_\rho}$  as well as $\bar{Y}_h(q^2,E_\rho)$ and $\Delta Y_h(q^2,E_\rho)$ are relegated to Appendix \ref{YY}.

If the $\rho$ helicity is not reconstructed, we sum the $L$ and $T$ sectors of Eq.~\eqref{YYEQ}. Then we have
\begin{align}
\label{boost}
    \frac{d^2\Gamma}{dq^2dE_\rho}&=\frac{d^2\Gamma_L}{dq^2dE_\rho}+\frac{d^2\Gamma_T}{dq^2dE_\rho}\nonumber\\
    &=\frac{d\Gamma}{dq^2}\left(\frac{1}{E_\rho^{\text{max}}-E_\rho^{\text{min}}}+\alpha_\rho P_\tau^{D^*}(q^2)\frac{2E_\rho-E_\rho^{\text{max}}-E_\rho^{\text{min}}}{\left(E_\rho^{\text{max}}-E_\rho^{\text{min}}\right)^2}\right),
\end{align}
where $E_\rho^{\text{max}}$ and $E_\rho^{\text{min}}$ are the maximum and minimum values of $E_\rho$ allowed by the kinematics in the $W^*$ rest frame:
\begin{align}
    E_\rho^{\text{max}}=\frac{q^2+m_\rho^2}{2\sqrt{q^2}}, \quad E_\rho^{\text{min}}=\frac{m_\tau^4+m_\rho^2 q^2}{2m_\tau^2\sqrt{q^2}},
\end{align}
$\alpha_\rho=\frac{m_{\tau }^2-2 m_\rho^2}{m_{\tau }^2+2 m_\rho^2}$ is the spin-analyzing power of $\rho$. It is the sensitivity coefficient of the energy spectrum of $\rho$ with respect to the polarization direction of $\tau$, and numerically, $\alpha_\rho\approx0.449$. If the final state is the $a_1$ particle, then the spin-analyzing power $\alpha_{a_1}\approx0.021$, which is much smaller than $\alpha_\rho$.
We can observe that $\frac{d^2\Gamma}{dq^2dE_\rho}$ is linear with respect to $E_\rho$. This enables us to easily extract $P_{\tau}^{D^*}$ from the energy spectrum of $\rho$. It should be noted again that the linear form is strictly valid only for the $L+T$ total spectrum. However, the individual $L$ and $T$ components are not in a linear form in the $W^*$ rest frame, and the exact shape is given by $\bar{Y}_h$ and $\Delta Y_h$ in Eq.~\eqref{YYEQ}.

If we boost Eq.~\eqref{boost} from the $W^*$ rest frame along the flight direction of $\tau$ to the $\tau$ rest frame, we can obtain
\begin{align}
\label{theta'}
    \frac{d^2\Gamma}{dq^2d\cos \theta'}=\frac{1}{2}\frac{d\Gamma}{dq^2}\left(1+\alpha_\rho P_\tau^{D^*}\cos\theta'\right),
\end{align}
where $\theta'$ is the polar angle of the $\rho$ motion direction in the $\tau$ rest frame and
\begin{align}
    \cos\theta'=\frac{2E_\rho-E_\rho^{\text{max}}-E_\rho^{\text{min}}}{E_\rho^{\text{max}}-E_\rho^{\text{min}}}.
\end{align}
Equation \eqref{theta'} reproduces very well the
result in Ref.~\cite{Hagiwara:1989fn} by Hagiwara, Martin and Zeppenfeld: for $\tau\rightarrow V\nu_\tau$ decays, $\left(1/\Gamma\right)d\Gamma/d\cos\theta'=\frac{1}{2}\left(1+\alpha_V P_\tau \cos \theta' \right)$, with  $\alpha_\rho\approx0.449$
It is worth noting that Eq.~\eqref{theta'} in this paper was independently derived starting from the complete five-dimensional decay formalism. The agreement can be regarded as a double check of our derivations. Furthermore, the $P_\tau^{D^*}$ for the $\tau\to V \nu_\tau$ decay is a number (no $q^2$ dependence), while here it is a $q^2$-dependent function which is determined by the helicity structure of the production process $\bar{B}\rightarrow D^* \tau^-\bar \nu_\tau$.

On the other hand, if we distinguish the $L$ and $T$ components of $\rho$, then Eq.~\eqref{YYEQ} can provide two complementary extractions of $P_\tau^{D^*}(q^2)$.

\begin{figure}
\centering
\includegraphics[width=0.42\linewidth]{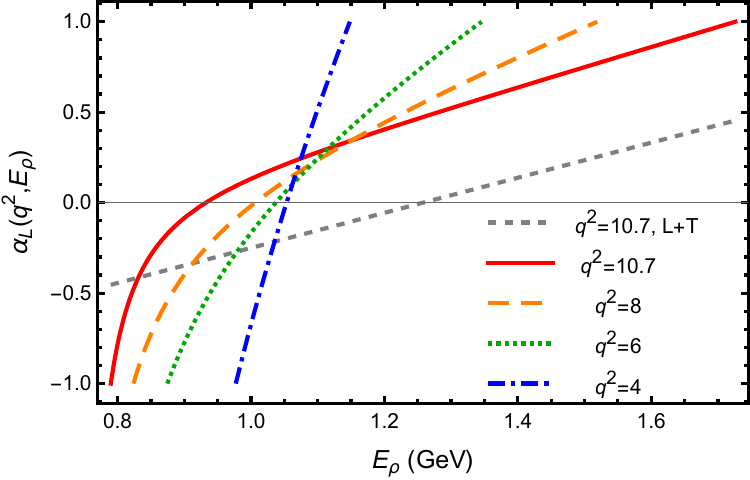}
\includegraphics[width=0.42\linewidth]{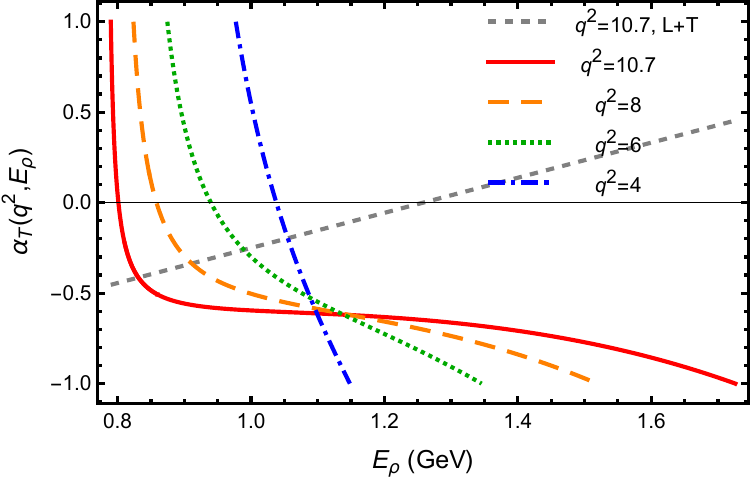}
\caption{Plot of $\alpha_L(q^2, E_\rho)$ and $\alpha_T(q^2, E_\rho)$ with $q^2=4, 6, 8, 10.7\, \text{GeV}^2$ (the maximum value of $q^2$). At the spectrum endpoints for all $q^2$, $\alpha_{L, T}=\pm1$. The gray dashed line corresponds to the helicity-summed case $\alpha_{L+T}=\frac{\Delta Y_L+\Delta Y_T}{\bar{Y}_L+\bar{Y}_T}$ with endpoint values $\pm\alpha_\rho=\pm0.449$.}
\label{alpha}
\end{figure}

The longitudinal and transverse components of the $\rho$ could be accessed experimentally through the $\rho\rightarrow\pi^+\pi^-$ decay. In the $\rho$ rest frame, with $\psi$ the $\pi$ helicity angle relative to the $\rho$ flight direction, the longitudinal and transverse angular distribution will be given by $\frac{d\Gamma_L}{d\cos\psi}=(3/2)\Gamma_L\cos^2\psi$ and $\frac{d\Gamma_T}{d\cos\psi}=(3/4)\Gamma_T\sin^2\psi$, respectively.
The effective analyzing powers will be defined by the coefficients $\bar Y_h(q^2, E_\rho)$ and $\Delta Y_h(q^2,E_\rho)$. That is,
\begin{align}
    \frac{d^2\Gamma_h}{dq^2dE_\rho}=\frac{d\Gamma}{dq^2}\bar{Y}_h(q^2, E_\rho)(1+\alpha_h(q^2, E_\rho)P_\tau^{D^*}(q^2)),
\end{align}
with
\begin{align}    \alpha_h(q^2,E_\rho)\equiv\frac{\Delta Y_h(q^2,E_\rho)}{\bar{Y}_h(q^2,E_\rho)}=\frac{d\Gamma(\tau_\uparrow\rightarrow\rho_h\nu_\tau)/dE_\rho-d\Gamma(\tau_\downarrow\rightarrow\rho_h\nu_\tau)/dE_\rho}{d\Gamma(\tau_\uparrow\rightarrow\rho_h\nu_\tau)/dE_\rho+d\Gamma(\tau_\downarrow\rightarrow\rho_h\nu_\tau)/dE_\rho}.
\end{align}
It measures the extent to which the $\rho$ energy spectrum responds to the polarization of $\tau$.
In Fig.~\ref{alpha}, we plot the curves showing how $\alpha_h$ changes with $E_\rho$ for different values of $q^2$.
At both ends of the spectrum,
\begin{align}
    \alpha_L(q^2,E_\rho^{\text{max}})=+1,\quad
    \alpha_L(q^2,E_\rho^{\text{min}})=-1,\nonumber\\
    \alpha_T(q^2,E_\rho^{\text{max}})=-1,\quad
    \alpha_T(q^2,E_\rho^{\text{min}})=+1.
\end{align}
The endpoint corresponds to $\theta'=0$ or $\pi$ in the $\tau$ rest frame. The endpoint analyzing powers $\pm1$ are exact for the individual $L$ or $T$ case, while the helicity-summed spectrum is everywhere bounded by $\alpha_\rho\approx0.449$ because of the cancellation between the two sectors. Note in Fig.~\ref{alpha}, $\alpha_{L+T}$ shows a line shape, not as in Eq.~\eqref{boost} where $\alpha_\rho$ is a constant of 0.449, due to different definitions of $\alpha$. In short, after separating the helicity of $\rho$, we obtain two additional quantities ($\alpha_L$ and $\alpha_T$) to test $P_\tau^{D^*}(q^2)$.

\section{Numerics and constraints on the new-physics couplings}\label{sec:numerical}
While the main results of the paper are contained in the previous sections,
 we provide here a brief discussion on the constraint of the NP couplings ($g_{L}, g_{R}, g_{P},g_{T}$) appearing in the effective Hamiltonian by fitting to the experimental value of $R_{D^*}$ and the $D^*$ longitudinal polarization fraction $\langle F_L^{D^*}\rangle$.  The allowed regions of the parameter space are shown graphically. We are not attempting a global fit to all the observables (as in Ref.~\cite{Murgui:2019czp}) in $\BDstartaunu$ decays which is beyond the scope of this paper.

 The masses of the particles are taken from PDG \cite{PDG}:
\begin{align}
        m_B=5.280 \,\text{GeV},\quad m_{D^*}=2.010 \operatorname{GeV},\quad
    m_{\tau}=1.777\operatorname{GeV},\nonumber\\
    m_b=4.183\operatorname{GeV},\quad
    m_c=1.273\operatorname{GeV},\quad
    m_\mu=0.1057\operatorname{GeV}.
\end{align}
The most recent SM and experimental values of $R_{D^*}$ are  \cite{HFLAV}
\begin{align}
    R_{D^*}^\text{SM}=0.258\pm0.005,\quad
    R_{D^*}^\text{exp}=0.281\pm0.011.
\end{align}
In this work, we obtain the SM prediction
as $R_{D^*}^\text{th}=0.253$ based on the HQET form factor values \cite{1309.0301}.

For $\langle F_{L}^{D^*}\rangle$, Belle has measured it to be $0.60\pm0.08\pm0.04$ \cite{Belle:2019ewo}, and we will adopt the most recent experimental value from LHCb \cite{LHCb:2023ssl} as
\begin{align}
    \langle F_{L}^{D^*}\rangle^{\text{exp}}=0.41\pm0.06\pm 0.03,
\end{align}
which is consistent with the SM prediction as well as the Belle result. The theoretical value in SM converges to a value in a range between 0.43
and 0.46 (See Ref.~\cite{Faustov:2022ybm} and references in \cite{Belle:2019ewo}).
The calculation of our theoretical values follows from
\begin{align}
    \langle F_{L}^{D^*}\rangle^{\text{th}}=\frac{\int F_L^{D^*} \frac{d\Gamma}{dq^2}d q^2}{\int\frac{d\Gamma}{dq^2}dq^2}=\frac{\int A(q^2)dq^2}{\int [A(q^2)+2B(q^2)]dq^2}.
\end{align}

\begin{figure}[htbp]
\centering
\includegraphics[width=0.45\linewidth]{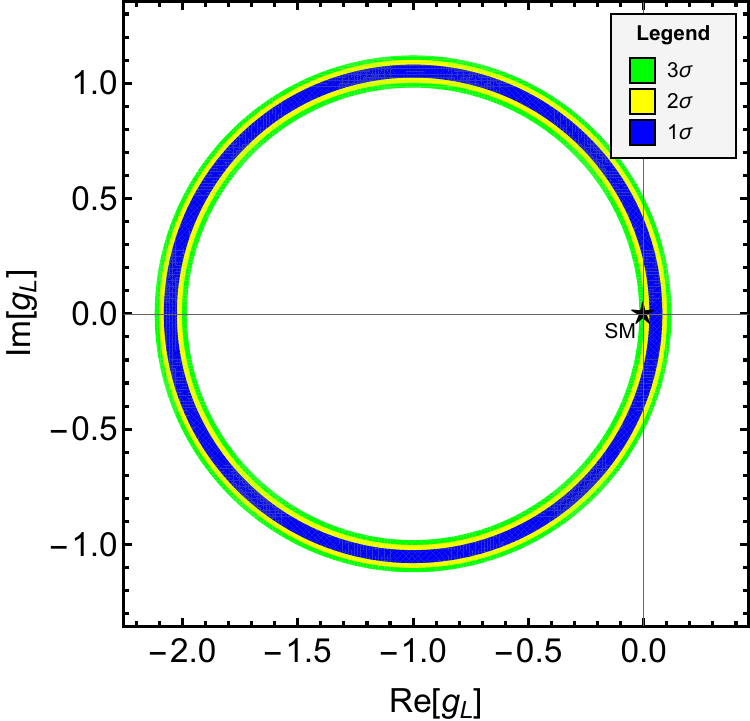}
\includegraphics[width=0.45\linewidth]{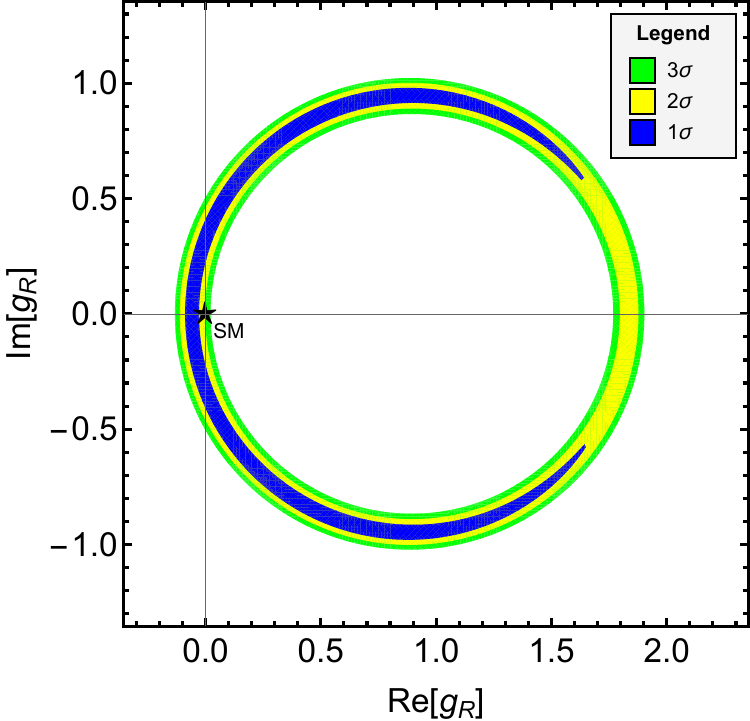}\\[4ex]
\includegraphics[width=0.4\linewidth]{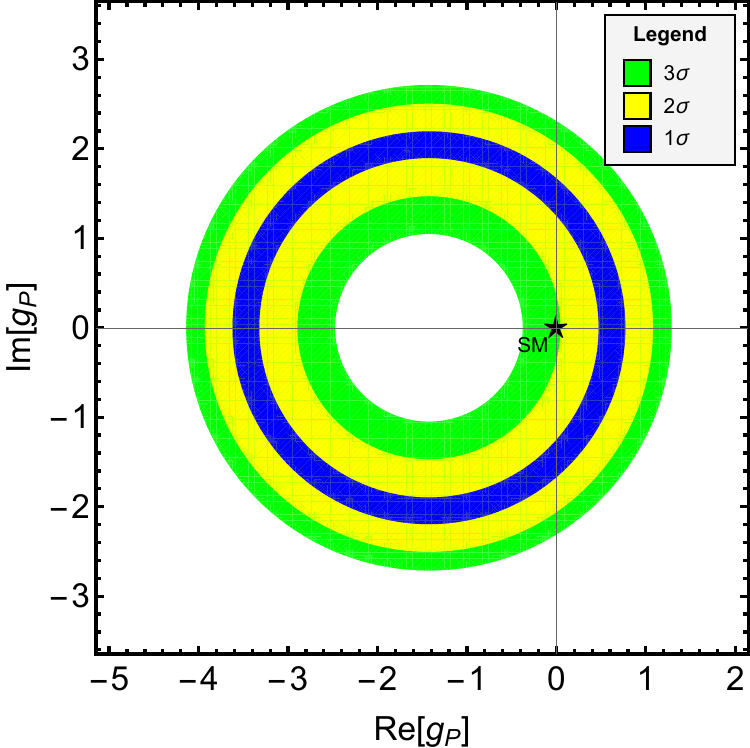}
\includegraphics[width=0.42\linewidth]{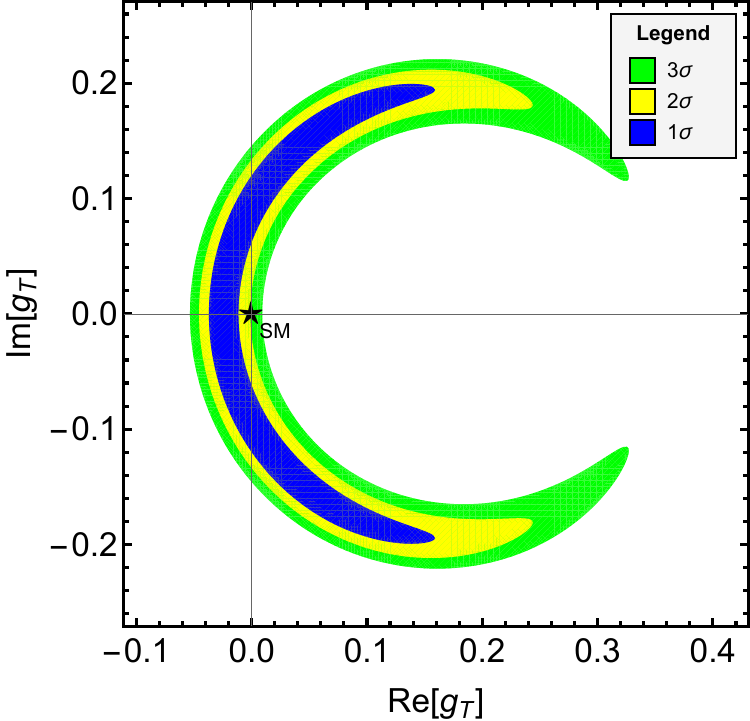}\\[4ex]
\caption{The allowed parameter space for $g_{i}$ considering the experimental values of $R_{D^*}^\text{exp}$ and $\langle F_L^{D^*}\rangle^{\text{exp}}$. The asterisk at the origin, $g_i=0$, indicates the location of the SM. The blue, yellow, and green band represent the $1\sigma, 2\sigma$ and $3\sigma$ levels of the experimental measurements, respectively.}
\label{gL}
\end{figure}

$R_{D^*}$ and $\langle F_{L}^{D^*}\rangle$ depend on the four complex-valued parameters $g_i$ with $i=L, R, P, T$. One can explore the correlation between the real and imaginary part of one $g_i$, assuming all others vanish. Such constraint plots are given in Fig.~\ref{gL}. The horizontal axis represents the real part, and the vertical axis represents the imaginary part. The areas in blue, yellow and green represent the ranges of $R_{D^*}^{\text{exp}}$ and $\langle F_{L}^{D^*}\rangle^{\text{exp}}$ at the $1\sigma$, $2\sigma$ and $3\sigma$, levels respectively. The asterisks at the origin (the location of SM) typically lie between the $2\sigma$ and $3\sigma$ regions. One observes that the current experimental value still leaves some room for accommodating non-zero values of $g_i$.

\begin{figure}[htbp]
\centering
\includegraphics[width=0.42\linewidth]{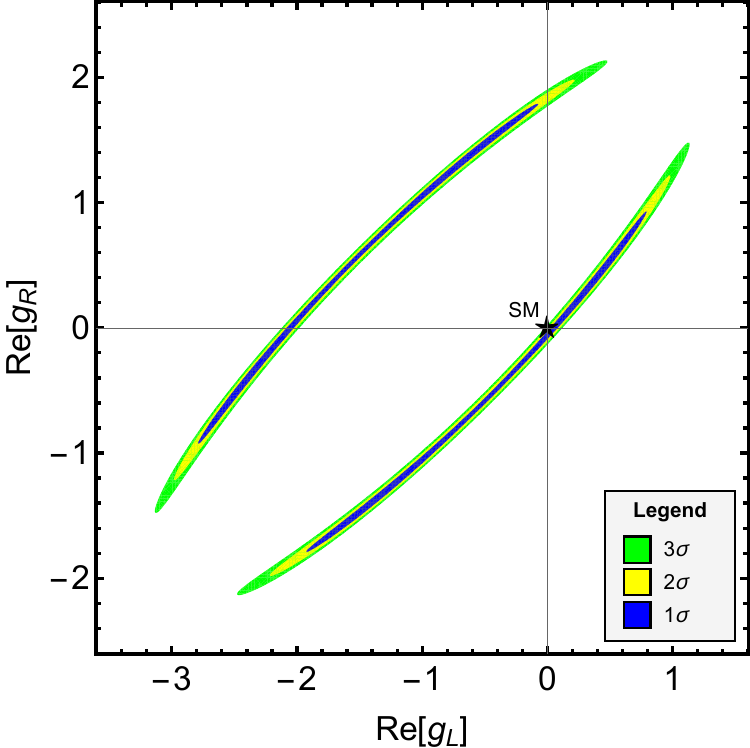}
\includegraphics[width=0.42\linewidth]{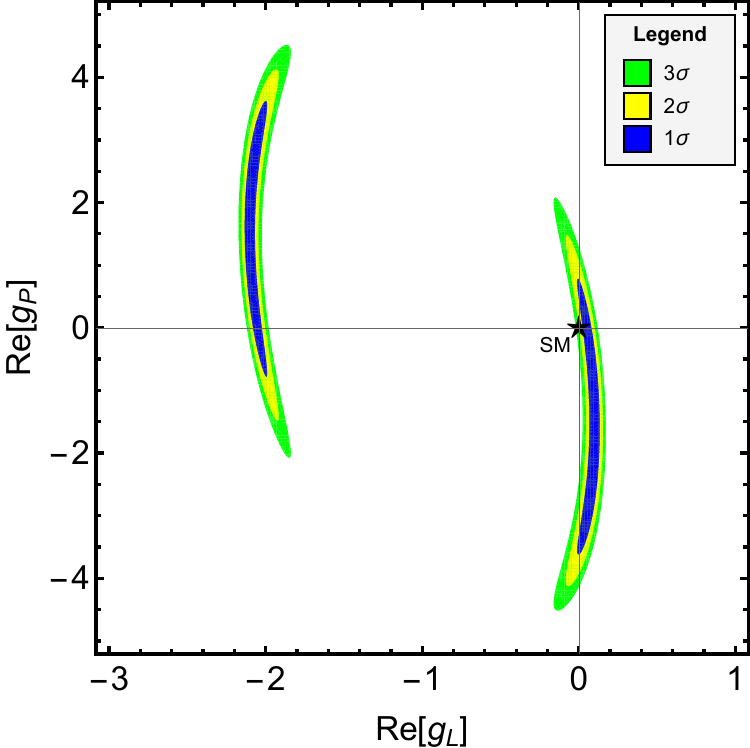}\\[4ex]
\includegraphics[width=0.44\linewidth]{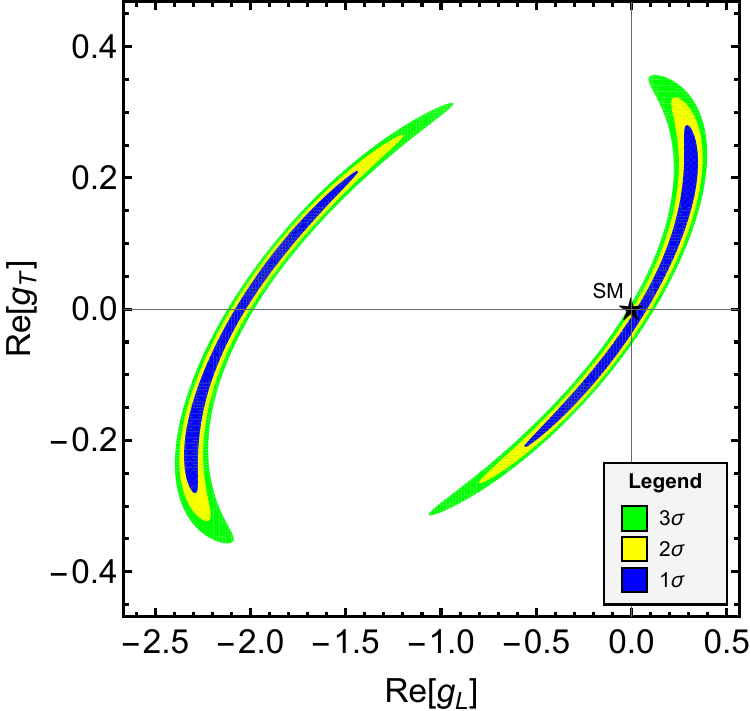}
\includegraphics[width=0.42\linewidth]{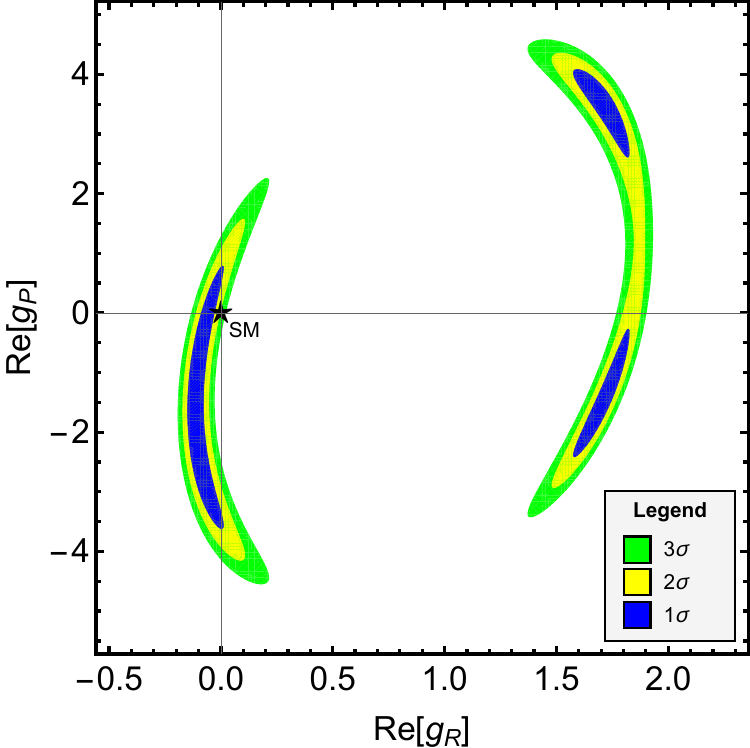}\\[4ex]
\includegraphics[width=0.44\linewidth]{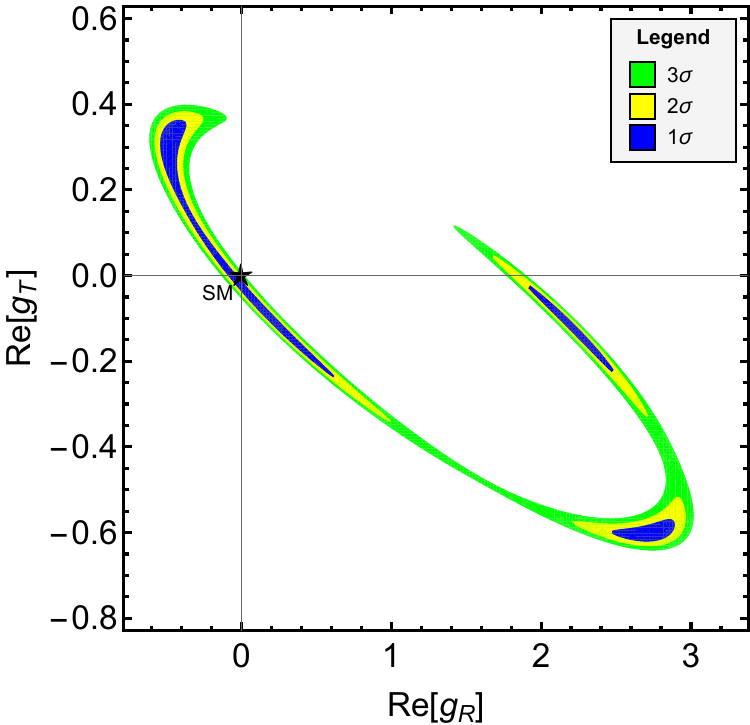}
\includegraphics[width=0.44\linewidth]{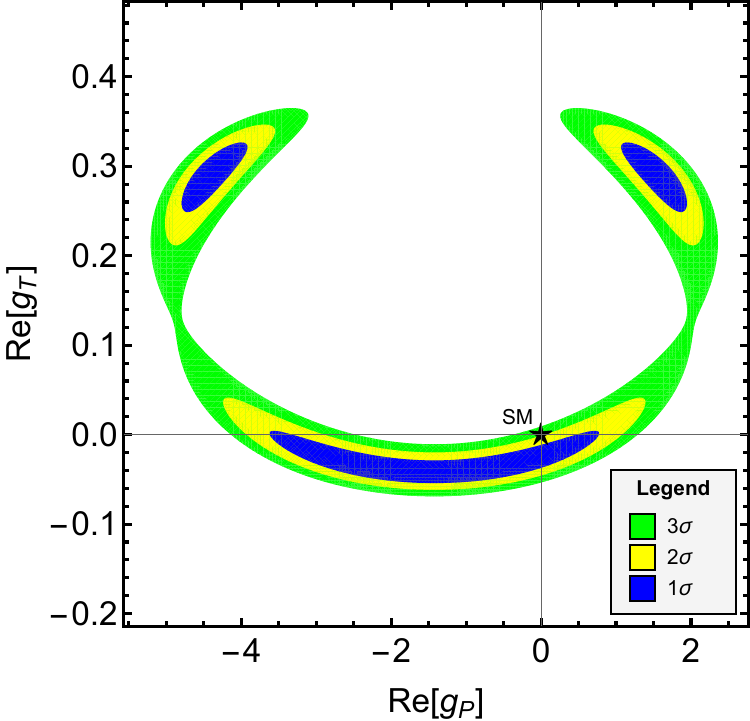}
\caption{The allowed parameter space for $\text{Re}[g_i]$ vs $\text{Re}[g_j]$. Same style as in Fig.~\ref{gL}.}
\label{gLgP}
\end{figure}

In addition to exploring the influence of the individual coupling coefficients, we can also investigate the combined effect of the two coupling coefficients.
Assuming that all imaginary parts of the coupling coefficients are zero, we plot the constraints of $(\text{Re}\, g_i, \text{Re}\, g_j)$ in Fig.~\ref{gLgP}.

We point out that
\begin{itemize}
\item The current $R_{D^*}^\text{exp}$ and  $\langle F_L^{D^*}\rangle^{\text{exp}}$ values do not rule out the SM. However,  improvement of their precision will narrow the allowed parameter space.
\item $R_{D^*}^\text{exp}$ and/or $\langle F_L^{D^*}\rangle^{\text{exp}}$
are not enough to completely determine the NP couplings. Different combinations of $g_i$ are possible and a full angular distribution analysis is required for distinguishing among the various NP couplings.
\end{itemize}

\section{Conclusion}\label{Conclusion}

In this work, we have performed a comprehensive analysis of the angular distribution in the decay chain $\bar{B}\rightarrow D^*(\rightarrow D\pi)\tau^-(\rightarrow V\nu_\tau)\bar{\nu}_\tau$ as a probe of new physics in the $b\rightarrow c\tau\nu$ transitions, where $V$ is a vector meson such as the $\rho$ or $a_1$ meson.

First, using the  decay $\tau^-\rightarrow\rho^-\nu_\tau$, we constructed a measurable five-dimensional angular distribution that avoids direct reconstruction of the $\tau$ momentum, thus eliminating systematic uncertainties associated with the modeling of the decay of the $\tau$. The final expression, given in Eq.~\eqref{angulardistribution}, depends on angular functions of the five kinematic variables $\theta^*$, $E_\rho$, $\theta_\rho$, $\chi_\rho$ and $q^2$ combined with the corresponding coefficient functions. This extends previous studies that considered the $\tau\rightarrow\pi\nu_\tau$ channel and offers a new and complementary path to investigating  possible new physics.  Based on this angular distribution, we constructed several integrated observables, such as the $D^*$ polarization fractions $F_L^{D^*}$ and $F_T^{D^*}$, the forward-backward asymmetry $\eta_1(E_\rho,q^2)$ and $\eta_2(E_\rho,q^2)$, which are experimentally measurable.

Second, by exploiting the helicity information of vector meson, we derived the expression for the spin analyzing power of the vector meson, related to the experimental extraction of the $\tau$ polarization $P_\tau^{D^*}(q^2)$ from the energy spectrum. With the helicity-summed spectrum, we found the analyzing power $\alpha_\rho \simeq 0.45$ and $\alpha_{a_1} \simeq 0.02$.  With the helicity-resolved case, we obtain the analyzing power $\alpha_h(q^2, E_\rho)$ with $h = L, T$. The expression for $P_\tau^{D^*}(q^2)$ and the numerical illustration under the Standard Model are given.

Using experimental values of $R_{D^*}^{\text{exp}}$ and $\langle F_L^{D^*}\rangle^{\text{exp}}$, we performed a numerical analysis to constrain the new-physics couplings $g_{L}$, $g_{R}$, $g_{P}$, and $g_{T}$ appearing in the effective Hamiltonian. In all cases, the SM point lies within the $2\sigma-3\sigma$ region, indicating that the current data does not exclude the SM, but allow a range of new-physics contributions. The future precise  measurement of $R_{D^*}$ and other observables will either point to the existence of new physics, or provide  stringent constraints on the new-physics couplings.

Finally, we strongly propose a complete experimental analysis of the angular observables presented in this work, which would allow for the extraction of new-physics parameters and/or the potential discovery of CP-violating observable  (as also a smoking-gun signal of new physics) in this channel, highlighting the power of angular analysis as a new physics analyzer in the era of precision flavor physics.

\section*{Acknowledgement}
The authors thank Drs. Saeed Kamali and Tejhas Kapoor for valuable discussions. The work of XWK is supported in part by the National Natural Science Foundation of China under Project No. 12275023. A.D. is supported in part by the U.S. National Science Foundation under Grant No.~PHY-2309937.

\begin{appendix}

\section{Phase Space}
The differential decay rate for $B\rightarrow D^* (\rightarrow D\pi) W^*[\rightarrow\tau(\rightarrow \rho \nu_\tau)\bar\nu_\tau]$ is written as
\begin{equation}
d\Gamma=\frac{1}{2m_B}|\mathcal{M}|^2d\Phi_n,
\end{equation}
where $\mathcal{M}$ is the decay amplitude encoding the dynamic transition and $d\Phi_n$ represents the final-state phase-space factor defined as
\begin{equation}
d\Phi_n=(2\pi)^4 \delta^4 (P-\sum_{i=1}^{n}p_i)\prod_{i=1}^{n}\frac{d^3{\bf p}_i}{(2\pi)^32E_i}.
\end{equation}
Here $P$ is the momentum of the parent particle, and $p_i$ and $E_i$ denote the four-momentum and energy of the final-state particle. For the five-body final state considered here,
\begin{align}
\Phi_5&=(2\pi)^4\int\delta^4(p_B-p_D-p_\pi-p_\rho-p_\nu-p_{\bar\nu})\nl
&\quad\times\frac{d^3{\bf p}_D}{(2\pi)^32E_D}\frac{d^3{\bf p}_\pi}{(2\pi)^32E_\pi}\frac{d^3{\bf p}_\rho}{(2\pi)^32E_\rho}\frac{d^3{\bf p}_\nu}{(2\pi)^32E_\nu}
\frac{d^3{\bf p}_{\bar\nu}}{(2\pi)^32E_{\bar\nu}}\nl
&=\frac{1}{2^{16}\pi^{11}}\int\delta^4(p_B-p_D-p_\pi-p_\rho-p_\nu-p_{\bar\nu})\frac{d^3{\bf p}_D}{E_D}\frac{d^3{\bf p}_\pi}{E_\pi}\frac{d^3{\bf p}_\rho}{E_\rho}\frac{d^3{\bf p}_\nu}{E_\nu}\frac{d^3{\bf p}_{\bar\nu}}{E_{\bar\nu}}\nl
&=\frac{1}{2^{16}\pi^{11}}\int d^4 p_{D^*}  d^4q \delta^4(p_B-p_{D^*}-q) \delta^4(p_{D^*}-p_D-p_\pi) \delta^4(q-p_\tau-p_{\bar\nu}) \nl &\quad\times\frac{d^3{\bf p}_D}{E_D}\frac{d^3{\bf p}_\pi}{E_\pi} \frac{d^3{\bf p}_\rho}{E_\rho}\frac{d^3{\bf p}_\nu}{E_\nu}\frac{d^3{\bf p}_{\bar\nu}}{E_{\bar\nu}}\nl
&=\frac{1}{2^{16}\pi^{11}}\int d^4 p_{D^*}d^4q d^4p_\tau \delta^4(p_B-p_{D^*}-q) \delta^4(p_{D^*}-p_D-p_\pi) \nl&\quad\times\delta^4(q-p_\tau-p_{\bar\nu})\delta^4(p_\tau-p_\rho-p_\nu)\frac{d^3{\bf p}_D}{E_D}\frac{d^3{\bf p}_\pi}{E_\pi}\frac{d^3{\bf p}_{\bar\nu}}{E_{\bar\nu}} \frac{d^3{\bf p}_\rho}{E_\rho}\frac{d^3{\bf p}_\nu}{E_\nu}.
\end{align}

For the off-shell momentum, we can insert the following relations:
\begin{align}
d^4 q&=d^3 {\bf q} dE_q=d^3{\bf q} d \sqrt{{\bf q}^2+q^2}=d^3{\bf q} \frac{d q^2}{2E_q},\nl
d^4 p_{D^*}&=d^3{\bf p}_{D^*} \frac{d p^2_{D^*}}{2E_{D^*}},\quad E_{D^*}=\sqrt{{\bf p}_{D^*}^2+p_{D^*}^2},\nl
d^4 p_{\tau}&=d^3{\bf p}_{\tau} \frac{d p^2_{\tau}}{2E_{\tau}}, \quad E_{\tau}=\sqrt{{\bf p}_\tau^2+p_\tau^2}.
 \end{align}
Note both sides of this equation are Lorentz invariant.
Then we have
\begin{align}\label{Phi5}
\Phi_5 &=\frac{1}{2^{19}\pi^{11}}\int \frac{d^3 {\bf p}_{D^*}}{E_{D^*}}d p_{D^*}^2 \frac{d^3{\bf q}}{E_q}dq^2  \frac{d^3{\bf p}_\tau}{E_\tau}d p_\tau^2
\delta^4(p_B-p_{D^*}-q)\delta^4(p_{D^*}-p_D-p_\pi)\nl&\times\delta^4(q-p_\tau-p_{\bar\nu}) \delta^4(p_\tau-p_\rho-p_\nu)
\frac{d^3{\bf p}_D}{E_D}\frac{d^3{\bf p}_\pi}{E_\pi}\frac{d^3{\bf p}_{\bar\nu}}{E_{\bar\nu}}\frac{d^3{\bf p}_\rho}{E_\rho}\frac{d^3{\bf p}_\nu}{E_\nu}
\end{align}

We split Eq.~\eqref{Phi5} into three parts, namely $I_1$, $I_2$, and $I_3$, for calculation.
The calculation for $B\rightarrow D^*W^*$part, $I_1$, is done in the $B$ rest frame:
\begin{align}
I_1=&\int\delta^4\left(p_B-p_{D^*}-q\right)\frac{d^3 {\bf p}_{D^*}}{E_{D^*}}\frac{d^3{\bf q}}{E_q}d p_{D^*}^2 dq^2\nl
=&\int\delta\left(E_B-E_{D^*}-E_q\right)\delta^3\left({\bf p}_{D^*}+{\bf q}\right)\frac{d^3 {\bf p}_{D^*}}{E_{D^*}}\frac{d^3{\bf q}}{E_q}d p_{D^*}^2 dq^2\nl
=&\int\delta\left(m_B-\sqrt{p_{D^*}^2+{\bf p}_{D^*}^2}-\sqrt{q^2+{\bf p}_{D^*}^2}\right)\frac{d^3{\bf p}_{D^*}}{E_{D^*}E_q}d p_{D^*}^2 dq^2\nl
=&\int\delta\left(|{\bf p}_{D^*}|-\frac{\sqrt{(m_B^2-m_{D^*}^2-q^2)^2-4m_{D^*}^2q^2}}{2m_B}\right)\frac{|{\bf p}_{D^*}| d|{\bf p}_{D^*}|}{E_{D^*}+E_q}d\Omega_{D^*} d p_{D^*}^2 dq^2\nl
=&4\pi\int\delta\left(|{\bf p}_{D^*}|-\frac{\sqrt{(m_B^2-m_{D^*}^2-q^2)^2-4m_{D^*}^2q^2}}{2m_B}\right)\frac{|{\bf p}_{D^*}| d|{\bf p}_{D^*}|}{E_{D^*}+E_q}d p_{D^*}^2 dq^2\nl
=&4\pi\int\frac{|{\bf p}_{D^*}|}{m_B}d p_{D^*}^2 dq^2.
\end{align}
The solid angle $d \Omega_{D^*}$ in $B$ rest frame trivially gives a factor of  $4\pi$.

$I_2$ for the $D^*\rightarrow D\pi$ part is calculated in the $D^*$ rest frame:
\begin{align}
 I_2=&\int\delta^4\left(p_{D^*}-p_D-p_\pi\right)\frac{d^3{\bf p}_D}{E_D}\frac{d^3{\bf p}_\pi}{E_\pi}\nl
=&\int\delta\left(E_{D^*}-E_D-E_\pi\right)\delta^3\left({\bf p}_D+{\bf p}_\pi\right)\frac{d^3{\bf p}_D}{E_D}\frac{d^3{\bf p}_\pi}{E_\pi}\nl
=&\int\delta\left(E_{D^*}-E_D-E_\pi\right)\frac{d^3{\bf p}_D}{E_D E_\pi}\nl
=&\int\delta\left(m_{D^*}-\sqrt{{\bf p}_D^2+m_D^2}-\sqrt{{\bf p}_D^2+m_\pi^2}\right)\frac{{\bf p}_D^2d|{\bf p}_D|}{E_D E_\pi}d\cos\theta^*d\chi^*\nl
=&2\pi\int\delta\left(|{\bf p}_D|-\frac{\sqrt{(m_{D^*}^2-m_D^2-m_\pi^2)^2-4m_D^2m_\pi^2}
}{2E_{D^*}}\right)\frac{|{\bf p}_D|d|{\bf p}_D|}{E_\pi+E_D}d\cos\theta^*\nl
=&2\pi\int\frac{|{\bf p}_D|}{m_{D^*}}d\cos\theta^*.
\end{align}
As usual, the integration over the azimuthal angle $\chi^*$ of $D$ in the $D^*$ rest frame gives a factor of $2\pi$.

As mentioned, ${\bf p}_\tau$ can not be measured, and so the kinematic variables of  $W^*\rightarrow\tau\bar\nu$ and $\tau\rightarrow\rho\nu$ parts are expressed in the $W^*$ rest frame. The integration, $I_3$, is written as
\begin{align}
I_3=&\int\delta^4\left(q-p_\tau-p_{\bar\nu}\right)\frac{d^3{\bf p}_\tau}{E_\tau}\frac{d^3{\bf p}_{\bar\nu}}{E_{\bar\nu}}
\delta^4\left(p_\tau-p_\rho-p_\nu\right)\frac{d^3{\bf p}_\rho}{E_\rho}\frac{d^3{\bf p}_\nu}{E_\nu}d p_\tau^2\nl
=&
\int\delta\left(E_q-E_\tau-E_{\bar\nu}\right)\delta^3\left({\bf q}-{\bf p}_\tau-{\bf p}_{\bar\nu}\right)\frac{d^3{\bf p}_\tau}{E_\tau}\frac{d^3{\bf p}_{\bar\nu}}{E_{\bar\nu}}\nl
&\times\delta(E_\tau-E_\rho-E_\nu)\delta^3({\bf p}_\tau-{\bf p}_\rho-{\bf p}_\nu)
\frac{d^3{\bf p}_\rho}{E_\rho}\frac{d^3{\bf p}_\nu}{E_\nu}d p_\tau^2\nl
=&
\int\delta\left(\sqrt{q^2}-E_\tau-E_{\bar\nu}\right)\delta^3\left({\bf p}_\tau+{\bf p}_{\bar\nu}\right)
\frac{d^3{\bf p}_\tau}{E_\tau}\frac{d^3{\bf p}_{\bar\nu}}{E_{\bar\nu}}\nl
&\times\delta\left(E_\tau-E_\rho-E_\nu\right)\delta^3\left({\bf p}_\tau-{\bf p}_\rho-{\bf p}_\nu\right)
\frac{d^3{\bf p}_\rho}{E_\rho}\frac{d^3{\bf p}_\nu}{E_\nu}d p_\tau^2\nl
=&
\int\delta\left(\sqrt{q^2}-E_\tau-E_{\bar\nu}\right)
\frac{d^3{\bf p}_\tau}{E_\tau E_{\bar\nu}}
\delta\left(E_\tau-E_\rho-E_\nu\right)
\frac{d^3{\bf p}_\rho}{E_\rho E_\nu}d p_\tau^2\nl
=& \int\delta\left(\sqrt{q^2}-\sqrt{{\bf p}_\tau^2+m_\tau^2}-|{\bf p}_\tau|\right)\frac{{\bf p}_\tau^2 d|{\bf p}_\tau|}{E_\tau E_{\bar\nu}}d\cos\theta_{\tau\rho}d\chi_{\tau\rho}\nl
&\times\delta\left(E_\tau-E_\rho-\sqrt{{\bf p}_\tau^2 + {\bf p}_\rho^2 - 2|{\bf p}_\tau||{\bf p}_\rho|\cos\theta_{\tau\rho}}\right)\frac{{\bf p}_\rho^2d|{\bf p}_\rho|}{E_\rho E_\nu}d\cos\theta_\rho d\chi_\rho d p_\tau^2\nl
=&\int\delta\left(|{\bf p}_\tau|-\frac{q^2-m_\tau^2}{2\sqrt{q^2}}\right)\frac{|{\bf p}_\tau|d|{\bf p}_\tau|}{E_\tau+E_{\bar{\nu}}}d\cos\theta_{\tau\rho}d\chi_{\tau\rho}\nl
&\times\delta\left(\cos\theta_{\tau\rho} - \frac{2E_\tau E_\rho-m_\tau^2-m_\rho^2}{2|{\bf p}_\tau||{\bf p}_\rho|}\right)\frac{d E_\rho}{|{\bf p}_\tau|}d\cos\theta_\rho d\chi_\rho d p_\tau^2\nl
=&\int\frac{1}{\sqrt{q^2}}dE_\rho d\cos\theta_\rho d\chi_\rho  d\chi_{\tau\rho}d p_\tau^2.
 \end{align}

Therefore, the five-body phase space is given by
\begin{equation}
\Phi_5=\frac{1}{2^{16}\pi^{9}}\frac{1}{\sqrt{q^2}}\frac{|{\bf p}_{D^*}||{\bf p}_D|}{m_Bm_{D^*}}\int d\cos\theta^*dE_\rho d\cos_\rho d\chi_\rho dq^2 d p_{D^*}^2 d p_\tau^2d\chi_{\tau\rho},
\end{equation}
and the differential decay rate is then
\begin{equation}
\frac{d^5\Gamma}{d\cos\theta^*dE_\rho d\cos\theta_\rho d\chi_\rho dq^2 }=\frac{1}{2^{17}\pi^{9}}\frac{1}{\sqrt{q^2}}\frac{|{\bf p}_{D^*}||{\bf p}_D|}{m_B^2m_{D^*}}\int |\mathcal{M}|^2d p_{D^*}^2 d p_\tau^2d\chi_{\tau\rho}.
\end{equation}

\section{Expressions of  $\mathcal{C}_i^S$, $\mathcal{C}_{ij}^R$ and $\mathcal{C}_{ij}^I$ in Eq.~\eqref{eq:d5Gamma_v1}}
\label{amplitudecoefficient}

The expressions for $\mathcal{C}_i^S$, $\mathcal{C}_{ij}^R$ and $\mathcal{C}_{ij}^I$ appearing in Eq.~\eqref{eq:d5Gamma_v1} are organized in Tables \ref{square term}, \ref{realpart of the cross term} and \ref{imaginarypart of the cross term}.

\begin{table}[h]
\caption{The terms $|\mathcal{A}_i|^2$ and their corresponding coefficients $\mathcal{C}_i^S$ in Eq.~\eqref{eq:d5Gamma_v1}. }\label{square term}
\centering
\begin{tabular}{|c|c|}
\hline
 $|\mathcal{A}_i|^2$ & Coefficients $\mathcal{C}_i^S$ \\
\hline
$|\mathcal{A}_{SP}|^2$ & $\mathcal{C}_{SP}^S=\mathcal{S}_{SP}\cos^2{\theta^*}$ \\
\hline
$|\mathcal{A}_{t}|^2$ & $\mathcal{C}_{t}^S=\mathcal{S}_{t}\cos^2\theta^*$ \\
\hline
$|\mathcal{A}_{0}|^2$ &
$\mathcal{C}_{0}^S=\left(\mathcal{S}_{0,1}+\mathcal{S}_{0,2}\cos2\theta_{\rho}\right)\cos^2{\theta^*}$ \\
\hline
$|\mathcal{A}_{\parallel}|^2$ &
$\mathcal{C}_{\parallel}^S=\begin{aligned}
\left(\mathcal{S}_{\parallel,1}+\mathcal{S}_{\parallel,2}\left(\cos {2\theta_\rho}-2\cos{2\chi_\rho}\sin^2{\theta_\rho}\right)\right)\sin^2\theta^*
\end{aligned}$
 \\
\hline
$|\mathcal{A}_{\perp}|^2$ &
$\begin{aligned}
\mathcal{C}_{\perp}^S=\left(\mathcal{S}_{\perp,1}+\mathcal{S}_{\perp,2}\left(\cos {2\theta_\rho}+2\cos{2\chi_\rho}\sin^2{\theta_\rho}\right)\right)\sin^2\theta^*
\end{aligned}$ \\
\hline
$|\mathcal{A}_{0 T}|^2$ & $\mathcal{C}_{0T}^S=\left(\mathcal{S}_{0T,1}+\mathcal{S}_{0T,2}\cos 2\theta_\rho\right)\cos^2\theta^*$ \\
\hline
$|\mathcal{A}_{\parallel T}|^2$ &
$\begin{aligned}
\mathcal{C}_{\parallel T}^S=\left(\mathcal{S}_{\parallel T,1}+\mathcal{S}_{\parallel T,2}\left(\cos {2\theta_\rho}-2\cos{2\chi_\rho}\sin^2{\theta_\rho}\right)\right)\sin^2\theta^*
\end{aligned}$ \\
\hline
 $ |\mathcal{A}_{\perp T}|^2 $ &
$\begin{aligned}
\mathcal{C}_{\perp T}^S=\left(\mathcal{S}_{\perp T,1}+\mathcal{S}_{\perp T,2}\left(\cos {2\theta_\rho}+2\cos{2\chi_\rho}\sin^2{\theta_\rho}\right)\right)\sin^2\theta^*
\end{aligned}$ \\
\hline
\end{tabular}
\end{table}

\begin{table}[h]
\caption{The terms $\operatorname{Re}[\mathcal{A}_i\mathcal{A}_j^*]$ and their corresponding coefficients $\mathcal{C}_{ij}^R$ in Eq.~\eqref{eq:d5Gamma_v1}.}
\label{realpart of the cross term}
\centering
\begin{tabular}{|c|c|}
\hline
 $\operatorname{Re}[\mathcal{A}_i\mathcal{A}_j^*]$&  Coefficients $\mathcal{C}_{ij}^R$\\
\hline
$\operatorname{Re}\big[\mathcal{A}_{t}\mathcal{A}_{0}^*\big]$& $\mathcal{C}_{t0}^R=\mathcal{R}_{t0}\cos{\theta_\rho}\cos^2{\theta^*}$ \\
\hline
$\operatorname{Re}\big[\mathcal{A}_{t}\mathcal{A}_{\parallel}^*\big]$ & $\mathcal{C}_{t\parallel}^R=\mathcal{R}_{t\parallel}\cos{\chi_\rho}\sin{\theta_\rho}\sin{2\theta^*}$\\
\hline
$\operatorname{Re}\big[\mathcal{A}_{0}\mathcal{A}_{\parallel}^*\big]$ & $\mathcal{C}_{0\parallel}^R=\mathcal{R}_{0\parallel}\cos{\chi_\rho}\sin{2\theta_\rho}\sin{2\theta^*}$\\
\hline
$\operatorname{Re}\big[\mathcal{A}_{0}\mathcal{A}_{\perp}^*\big]$ & $\mathcal{C}_{0\perp}^R=\mathcal{R}_{0\perp}\cos{\chi}_{\rho}\sin{\theta_\rho}\sin{2\theta^*}$\\
\hline
$\operatorname{Re}\big[\mathcal{A}_{\parallel}\mathcal{A}_{\perp}^*\big]$ & $\mathcal{C}_{\parallel\perp}^R=\mathcal{R}_{\parallel\perp}\cos{\theta_\rho}\sin^2{\theta^*}$\\
\hline
$
\operatorname{Re}\big[\mathcal{A}_{0T}\mathcal{A}_{\parallel T}^*\big]
$&
$\mathcal{C}_{0T\parallel T}^R=\mathcal{R}_{0T\parallel T}\cos\chi_\rho \sin2\theta_\rho \sin2\theta^*$\\
\hline

$
\operatorname{Re}\big[\mathcal{A}_{0T}\mathcal{A}_{\perp T}^*\big]
$&
$\mathcal{C}_{0T\perp T}^R=\mathcal{R}_{0T\perp T}\cos\chi_\rho\sin\theta_\rho\sin2\theta^*$\\
\hline

$
\operatorname{Re}\big[\mathcal{A}_{\parallel T}\mathcal{A}_{\perp T}^*\big]$ & $\mathcal{C}_{\parallel T\perp T}^R=\mathcal{R}_{\parallel T\perp T}\cos\theta_\rho\sin^2{\theta^*}$\\
\hline

$\operatorname{Re}\big[\mathcal{A}_{SP}\mathcal{A}_{t}^*\big]$ &
$\mathcal{C}_{SP t}^R=\mathcal{R}_{SPt}\cos^2{\theta^*}$\\
\hline
$\operatorname{Re}\big[\mathcal{A}_{SP}\mathcal{A}_{0}^*\big]$ & $\mathcal{C}_{SP0}^R=\mathcal{R}_{SP0}\cos\theta_\rho\cos^2{\theta^*}$\\
\hline
$\operatorname{Re}\big[\mathcal{A}_{SP}\mathcal{A}_{\parallel}^*\big]$ & $\mathcal{C}_{SP\parallel}^R=\mathcal{R}_{SP\parallel}\cos{\chi_\rho}\sin{\theta_\rho}\sin2\theta^*$\\
\hline

$\operatorname{Re}\big[\mathcal{A}_{SP}\mathcal{A}_{0T}^*\big]$
&
$\mathcal{C}_{SP0T}^R=\mathcal{R}_{SP0T}\cos\theta_\rho\cos^2\theta^*$\\
\hline

$\operatorname{Re}\big[\mathcal{A}_{SP}\mathcal{A}_{\parallel T}^*\big]$ & $\mathcal{C}_{SP\parallel T}^R=\mathcal{R}_{SP\parallel T}\cos\chi_\rho\sin{\theta_\rho}\sin2\theta^*$\\
\hline

$\operatorname{Re}\big[\mathcal{A}_{t}\mathcal{A}_{0T}^*\big]$&$\mathcal{C}_{t0T}^R=\mathcal{R}_{t0T}\cos{\theta_\rho}\cos^2\theta^*$\\
\hline

$\operatorname{Re}\big[\mathcal{A}_{t}\mathcal{A}_{\parallel T}^*\big]$&$\mathcal{C}_{t\parallel T}^R=\mathcal{R}_{t\parallel T}\cos\chi_{\rho} \sin\theta_\rho\sin2\theta^*$\\
\hline

$\operatorname{Re}\big[\mathcal{A}_{0}\mathcal{A}_{0 T}^*\big]$ & $\mathcal{C}_{00T}^R=\left(\mathcal{R}_{00T,1}+\mathcal{R}_{00T,2}\cos{2\theta_\rho}\right)\cos^2\theta^*$\\
\hline

$\operatorname{Re}\big[\mathcal{A}_{0}\mathcal{A}_{\parallel T}^*\big]$ & $\mathcal{C}_{0\parallel T}^R=\mathcal{R}_{0\parallel T}\cos\chi_\rho\sin{2\theta_\rho}\sin2\theta^*$\\
\hline
$\operatorname{Re}\big[\mathcal{A}_{0}\mathcal{A}_{\perp T}^*\big]$ & $\mathcal{C}_{0\perp T}^R=\mathcal{R}_{0 \perp T}\cos\chi_\rho\sin\theta_\rho\sin2\theta^*$\\
\hline

$\operatorname{Re}[\mathcal{A}_{\parallel}\mathcal{A}_{0T}^*]    $&$\mathcal{C}_{\parallel0T}^R=\mathcal{R}_{\parallel 0T}\cos\chi_\rho\sin2\theta_\rho\sin2\theta^*$\\
\hline

$\operatorname{Re}\big[\mathcal{A}_{\parallel}\mathcal{A}_{\parallel T}^*\big]$ & $\mathcal{C}_{\parallel\parallel T}^R=\begin{aligned}\left(\mathcal{R}_{\parallel\parallel T,1}+\mathcal{R}_{\parallel\parallel T,2}\left(\cos2\theta_\rho-2\cos2\chi_\rho\sin^2\theta_\rho\right)\right)\sin^2\theta^*\end{aligned}$\\
\hline
$\operatorname{Re}\big[\mathcal{A}_{\parallel}\mathcal{A}_{\perp T}^*\big]$ & $\mathcal{C}_{\parallel\perp T}^R=\begin{aligned}\mathcal{R}_{\parallel\perp T}\cos\theta_\rho \sin^2\theta^*\end{aligned}$\\
\hline

$\operatorname{Re}[\mathcal{A}_{\perp}\mathcal{A}_{0T}^*]$&$\mathcal{C}_{\perp0T}^R=\mathcal{R}_{\perp0T}\cos\chi_\rho\sin\theta_\rho\sin2\theta^*$\\
\hline

$\operatorname{Re}\big[\mathcal{A}_{\perp}\mathcal{A}_{\parallel T}^*\big]$ &
$\mathcal{C}_{\perp\parallel T}^R=\begin{aligned}\mathcal{R}_{\perp\parallel T}\cos\theta_\rho \sin^2\theta^*\end{aligned}$\\
\hline
$\operatorname{Re}\big[\mathcal{A}_{\perp}\mathcal{A}_{\perp T}^*\big]$ & $\mathcal{C}_{\perp\perp T}^R=\begin{aligned}\left(\mathcal{R}_{\perp\perp T,1}+\mathcal{R}_{\perp\perp T,2}\left(\cos2\theta_\rho+2\cos2\chi_\rho\sin^2\theta_\rho\right)\right)\sin^2\theta^*\end{aligned}$\\
\hline
\end{tabular}
\end{table}

\begin{table}[h]
\centering
\caption{The terms  $\operatorname{Im}[\mathcal{A}_{i}\mathcal{A}_j^*]$ and their corresponding coefficients $\mathcal{C}_{ij}^I$ in Eq.~\eqref{eq:d5Gamma_v1}. }
\label{imaginarypart of the cross term}
\begin{tabular}{|c|c|}
\hline
$\operatorname{Im}[\mathcal{A}_{i}\mathcal{A}_j^*]$ & Coefficients $\mathcal{C}_{ij}^I$ \\
\hline
$\operatorname{Im}\big[\mathcal{A}_{t}\mathcal{A}_{\perp}^*\big]$ & $\mathcal{C}_{t\perp}^I=\mathcal{I}_{t\perp}\sin\chi_\rho\sin\theta_\rho\sin2\theta^*$\\
\hline
$\operatorname{Im}\big[\mathcal{A}_{0}\mathcal{A}_{\perp}^*\big]$ & $\mathcal{C}_{0\perp}^I=\mathcal{I}_{0\perp}\sin\chi_\rho\sin2\theta_\rho\sin2\theta^*$\\
\hline
$\operatorname{Im}\big[\mathcal{A}_{\parallel}\mathcal{A}_{\perp}^*\big]$&
$\mathcal{C}_{\parallel\perp}^I=\mathcal{I}_{\parallel\perp}\sin2\chi_\rho\sin^2\theta_\rho\sin^2\theta^*$\\
\hline

$\operatorname{Im}\big[\mathcal{A}_{SP}\mathcal{A}_{\perp}^*\big]$ & $\mathcal{C}_{SP\perp}^I=\mathcal{I}_{SP\perp}\sin\chi_\rho\sin\theta_\rho\sin2\theta^*$\\
\hline

$\operatorname{Im}\big[\mathcal{A}_{SP}\mathcal{A}_{\perp T}^*\big]$ & $\mathcal{C}_{SP\perp T}^I=\mathcal{I}_{SP\perp T}\sin\chi_\rho\sin\theta_\rho\sin2\theta^*$\\
\hline

$\operatorname{Im}\big[\mathcal{A}_{t}\mathcal{A}_{\perp T}^*\big]$&$\mathcal{C}_{t\perp T}^I=\mathcal{I}_{t\perp T}\sin\chi_\rho\sin\theta_\rho\sin2\theta^*$\\
\hline
$\operatorname{Im}\big[\mathcal{A}_{0}\mathcal{A}_{\parallel T}^*\big]$&$\mathcal{C}_{0\parallel T}^I=\mathcal{I}_{0\parallel T}\sin\chi_\rho\sin\theta_\rho\sin2\theta^*$\\
\hline
$\operatorname{Im}\big[\mathcal{A}_{0}\mathcal{A}_{\perp T}^*\big]$&$\mathcal{C}_{0\perp T}^I=\mathcal{I}_{0\perp T}\sin\chi_\rho\sin2\theta_\rho\sin2\theta^*$\\
\hline

$\operatorname{Im}\big[\mathcal{A}_{\parallel}\mathcal{A}_{0 T}^*\big]$&$\mathcal{C}_{\parallel0T}^I=\mathcal{I}_{\parallel 0T}\sin\chi_\rho\sin\theta_\rho\sin2\theta^*$\\
\hline

$\operatorname{Im}\big[\mathcal{A}_{\parallel}\mathcal{A}_{\perp T}^*\big]$ & $\mathcal{C}_{\parallel\perp T}^I=\mathcal{I}_{\parallel\perp T}\sin2\chi_\rho\sin^2\theta_\rho\sin^2\theta^*
$\\
\hline

$\operatorname{Im}\big[\mathcal{A}_{\perp}\mathcal{A}_{0 T}^*\big]$&$\mathcal{C}_{\perp0T}^I=\mathcal{I}_{\perp0 T}\sin\chi_\rho\sin2\theta_\rho\sin2\theta^*$\\
\hline

$\operatorname{Im}\big[\mathcal{A}_{\perp}\mathcal{A}_{\parallel T}^*\big]$ & $\mathcal{C}_{\perp\parallel T}^R=\mathcal{I}_{\perp\parallel T}\sin2\chi_\rho\sin^2\theta_\rho\sin^2\theta^*
$\\
\hline
\end{tabular}
\end{table}

With the abbreviations
\begin{equation}
        \kappa_{\tau}=\frac{m_\tau}{\sqrt{q^2}},\quad
        \kappa_{\rho}=\frac{m_\rho}{\sqrt{q^2}},\quad
        \varepsilon_{\rho}=\frac{E_\rho}{\sqrt{q^2}},
\end{equation}
the expressions for the coefficients in Table \ref{square term} are
\begin{align}\label{SSP}
    \mathcal{S}_{SP}&=\frac{1}{\kappa _{\tau }^2}\mathcal{S}_{t}=16 \left(2 \varepsilon _{\rho } \kappa _{\tau }^2+\left(1-4 \varepsilon _{\rho }\right) \kappa _{\rho }^2+2 \kappa _{\rho }^4-\kappa _{\tau }^4\right),
    \\[2ex]
      \label{S01}
        \mathcal{S}_{0,1}&= \frac{4 }{\kappa _{\tau }^2 \left(\varepsilon _{\rho }^2-\kappa _{\rho }^2\right)}
        \bigg(4 \varepsilon _{\rho }^3 \kappa _{\tau }^2 \left(\kappa _{\tau }^2-1\right) \left(\kappa _{\tau }^2-2 \kappa _{\rho }^2\right)+2 \varepsilon _{\rho }^2 \left(\kappa _{\tau }^4+2 \kappa _{\tau }^2-1\right) \left(2 \kappa _{\rho }^4-\kappa _{\tau }^4\right)\nonumber\\
        &+2 \varepsilon _{\rho } \Big(-\kappa _{\rho }^2 \left(\kappa _{\tau }^2-3\right) \kappa _{\tau }^4+2 \kappa _{\rho }^4 \left(\kappa _{\tau }^2-3\right) \kappa _{\tau }^2-2 \kappa _{\rho }^6 \left(\kappa _{\tau }^2+1\right)+\kappa _{\tau }^8+\kappa _{\tau }^6\Big)\nonumber\\
        &+\left(2 \kappa _{\rho }^4-\kappa _{\tau }^4\right) \Big(-\kappa _{\rho }^2 \left(\kappa _{\tau }^4-3\right)+\kappa _{\rho }^4+\kappa _{\tau }^4\Big)\bigg),
  \\[2ex]
    \label{S02}
        \mathcal{S}_{0,2}&=-4\mathcal{S}_{\parallel,2}=-4\mathcal{S}_{\perp,2}=\frac{4}{\kappa _{\tau }^2 \left(\varepsilon _{\rho }^2-\kappa _{\rho }^2\right)}
        \bigg(4 \varepsilon _{\rho }^3 \left(\kappa _{\tau }^2+1\right) \kappa _{\tau }^2 \left(\kappa _{\tau }^2-2 \kappa _{\rho }^2\right)\nonumber\\
        &+2 \varepsilon _{\rho }^2 \Big(2 \kappa _{\rho }^4 \left(\kappa _{\tau }^4+6 \kappa _{\tau }^2+1\right)+2 \kappa _{\rho }^2 \kappa _{\tau }^4-\kappa _{\tau }^4 \left(\kappa _{\tau }^4+6 \kappa _{\tau }^2+1\right)\Big)\nonumber\\
        &+2 \varepsilon _{\rho } \left(\kappa _{\tau }^2+1\right) \left(-2 \kappa _{\rho }^4 \kappa _{\tau }^2+\kappa _{\rho }^2 \kappa _{\tau }^4-6 \kappa _{\rho }^6+3 \kappa _{\tau }^6\right)\nonumber\\
        &-\left(\kappa _{\rho }^2+3\right) \kappa _{\tau }^8+4 \kappa _{\rho }^2 \left(2 \kappa _{\rho }^4-\kappa _{\rho }^2-1\right) \kappa _{\tau }^4+8 \left(3 \kappa _{\rho }^8+\kappa _{\rho }^6\right)\bigg),
\\[2ex]
 \label{Spar1}
    \mathcal{S}_{\parallel,1}&=\mathcal{S}_{\perp,1}=\frac{1}{{\kappa _{\tau }^2 \left(\varepsilon _{\rho }^2-\kappa _{\rho }^2\right)}}
    \bigg(4 \varepsilon _{\rho }^3 \kappa _{\tau }^2 \left(\kappa _{\tau }^2-3\right) \left(\kappa _{\tau }^2-2 \kappa _{\rho }^2\right)\nonumber\\
    &+2 \varepsilon _{\rho }^2 \Big(-2 \kappa _{\rho }^2 \kappa _{\tau }^4+2 \kappa _{\rho }^4 \left(\kappa _{\tau }^4-2 \kappa _{\tau }^2-3\right)-\kappa _{\tau }^8+2 \kappa _{\tau }^6+3 \kappa _{\tau }^4\Big)\nonumber\\
    &+2 \varepsilon _{\rho } \Big(2 \kappa _{\rho }^6 \left(\kappa _{\tau }^2+1\right)+2 \kappa _{\rho }^4 \kappa _{\tau }^2 \left(3 \kappa _{\tau }^2-5\right)+\kappa _{\rho }^2 \left(5 \kappa _{\tau }^4-3 \kappa _{\tau }^6\right)-\kappa _{\tau }^6 \left(\kappa _{\tau }^2+1\right)\Big)\nonumber\\
    &+\kappa _{\rho }^6 \left(10-6 \kappa _{\tau }^4\right)+3 \kappa _{\rho }^4 \kappa _{\tau }^4+\kappa _{\rho }^2 \kappa _{\tau }^4 \left(3 \kappa _{\tau }^4-5\right)-2 \kappa _{\rho }^8+\kappa _{\tau }^8\bigg),
 \\[2ex]
   \label{S0T1}
    \mathcal{S}_{0T,1}&=\frac{64}{\varepsilon _{\rho }^2-\kappa _{\rho }^2} \bigg(4 \varepsilon _{\rho }^3 \left(\kappa _{\tau }^2-1\right) \left(2 \kappa _{\rho }^2-\kappa _{\tau }^2\right)-2 \varepsilon _{\rho }^2 \kappa _{\rho }^2 \left(\kappa _{\tau }^4-2 \kappa _{\tau }^2-1\right)\nonumber\\
    &+2 \varepsilon _{\rho } \kappa _{\rho }^2 \Big(\kappa _{\rho }^2 \left(3-5 \kappa _{\tau }^2\right)+\kappa _{\tau }^2 \left(\kappa _{\tau }^2-3\right)\Big)+\kappa _{\rho }^2 \Big(\kappa _{\rho }^2 \left(3 \kappa _{\tau }^4-1\right)+\kappa _{\rho }^4+\kappa _{\tau }^4\Big)\bigg),
 \\[2ex]
   \label{0T2}
    \mathcal{S}_{0T,2}&=-4 \mathcal{S}_{\parallel T,2}=-4 \mathcal{S}_{\perp T,2}\nonumber\\
    &=\frac{64}{\varepsilon _{\rho }^2-\kappa _{\rho }^2} \bigg(2 \varepsilon _{\rho } \kappa _{\rho }^2 \left(\kappa _{\tau }^2+1\right) \left(\kappa _{\rho }^2-5 \kappa _{\tau }^2\right)-4 \varepsilon _{\rho }^3 \left(\kappa _{\tau }^2+1\right) \left(2 \kappa _{\rho }^2-\kappa _{\tau }^2\right)\nonumber\\
    &+2 \varepsilon _{\rho }^2 \Big(\kappa _{\rho }^2 \left(\kappa _{\tau }^4+6 \kappa _{\tau }^2+1\right)+4 \kappa _{\rho }^4-2 \kappa _{\tau }^4\Big)+\kappa _{\rho }^4 \left(\kappa _{\tau }^4+1\right)+7 \kappa _{\rho }^2 \kappa _{\tau }^4-5 \kappa _{\rho }^6\bigg),
\\[2ex]
    \label{SparT1}
    \mathcal{S}_{\parallel T,1}&= \mathcal{S}_{\perp T,1}\nonumber\\
    &=\frac{16}{\varepsilon _{\rho }^2-\kappa _{\rho }^2} \bigg(-2 \varepsilon _{\rho } \kappa _{\rho }^2 \Big(\kappa _{\rho }^2 \left(11 \kappa _{\tau }^2-5\right)-7 \kappa _{\tau }^4+\kappa _{\tau }^2\Big)+4 \varepsilon _{\rho }^3 \left(3 \kappa _{\tau }^2-1\right) \left(2 \kappa _{\rho }^2-\kappa _{\tau }^2\right)\nonumber\\
    &+\varepsilon _{\rho }^2 \Big(\kappa _{\rho }^2 \left(-6 \kappa _{\tau }^4-4 \kappa _{\tau }^2+2\right)-8 \kappa _{\rho }^4+4 \kappa _{\tau }^4\Big)+\kappa _{\rho }^4 \left(5 \kappa _{\tau }^4-3\right)-5 \kappa _{\rho }^2 \kappa _{\tau }^4+7 \kappa _{\rho }^6\bigg).
\end{align}
The coefficients in Table~\ref{realpart of the cross term} are
\begin{align}\label{Rt0}
    \mathcal{R}_{t0}&=2 \sqrt{2} \mathcal{R}_{t\parallel}
    =\kappa _{\tau }\mathcal{R}_{SP0}=2 \sqrt{2} \kappa _{\tau }\mathcal{R}_{SP\parallel}=\frac{32}{\sqrt{\varepsilon _{\rho }^2-\kappa _{\rho }^2}}
    \bigg(2 \kappa _{\rho }^4 \Big(\varepsilon _{\rho } \left(\kappa _{\tau }^2+2\right)-1\Big)\nonumber\\
    &+\left(-4 \varepsilon _{\rho }^2+\varepsilon _{\rho }+1\right) \kappa _{\rho }^2 \kappa _{\tau }^2+\left(\varepsilon _{\rho }-1\right) \kappa _{\tau }^4 \left(2 \varepsilon _{\rho }-\kappa _{\tau }^2\right)-2 \kappa _{\rho }^6\bigg),
\\[2ex]
    \mathcal{R}_{0\parallel}&=\frac{2 \sqrt{2}}{\kappa _{\tau }^2 \left(\varepsilon _{\rho }^2-\kappa _{\rho }^2\right)}
    \bigg(4 \varepsilon _{\rho }^3 \left(\kappa _{\tau }^2+1\right) \kappa _{\tau }^2 \left(\kappa _{\tau }^2-2 \kappa _{\rho }^2\right)\nl&+2 \varepsilon _{\rho} \left(\kappa _{\tau }^2+1\right) \left(-2 \kappa _{\rho }^4 \kappa _{\tau }^2+\kappa _{\rho }^2 \kappa _{\tau }^4-6 \kappa _{\rho }^6+3 \kappa _{\tau }^6\right)\nonumber\\
    &+2 \varepsilon _{\rho }^2 \Big(2 \kappa _{\rho }^4 \left(\kappa _{\tau }^4+6 \kappa _{\tau }^2+1\right)+2 \kappa _{\rho }^2 \kappa _{\tau }^4-\kappa _{\tau }^4 \left(\kappa _{\tau }^4+6 \kappa _{\tau }^2+1\right)\Big)\nonumber\\
    &-\kappa _{\tau }^8\left(\kappa _{\rho }^2+3\right) +\kappa _{\rho }^2 \left(2 \kappa _{\rho }^4-\kappa _{\rho }^2-1\right) \kappa _{\tau }^4+2 \left(3 \kappa _{\rho }^8+\kappa _{\rho }^6\right)\bigg),
\\[2ex]
    \mathcal{R}_{0\perp}&=-\frac{1}{\sqrt{2}}\mathcal{R}_{\parallel\perp}=\frac{8 \sqrt{2} }{\kappa _{\tau }^2 \sqrt{\varepsilon _{\rho }^2-\kappa _{\rho }^2}}
    \bigg(\varepsilon _{\rho }^2 \left(4 \kappa _{\rho }^2 \kappa _{\tau }^2-2 \kappa _{\tau }^4\right)\nonumber\\
    &+\varepsilon _{\rho } \Big(-\kappa _{\rho }^2 \kappa _{\tau }^4-2 \kappa _{\rho }^4 \left(2 \kappa _{\tau }^2+1\right)+2 \kappa _{\tau }^6+\kappa _{\tau }^4\Big)+2 \kappa _{\rho }^4 \kappa _{\tau }^4-\kappa _{\rho }^2 \kappa _{\tau }^6+2 \kappa _{\rho }^6-\kappa _{\tau }^6\bigg),
\\[2ex]
    \mathcal{R}_{0T\parallel T}&=-\frac{1}{\sqrt{2}}\mathcal{R}_{\parallel T\perp T}\nonumber\\
    &=\frac{32 \sqrt{2} }{\varepsilon _{\rho }^2-\kappa _{\rho }^2}
    \bigg(2 \varepsilon _{\rho } \kappa _{\rho }^2 \left(\kappa _{\tau }^2+1\right) \left(\kappa _{\rho }^2-5 \kappa _{\tau }^2\right)-4 \varepsilon _{\rho }^3 \left(\kappa _{\tau }^2+1\right) \left(2 \kappa _{\rho }^2-\kappa _{\tau }^2\right)\nonumber\\
    &+2 \varepsilon _{\rho }^2 \Big(\kappa _{\rho }^2 \left(\kappa _{\tau }^4+6 \kappa _{\tau }^2+1\right)+4 \kappa _{\rho }^4-2 \kappa _{\tau }^4\Big)+\kappa _{\rho }^4 \left(\kappa _{\tau }^4+1\right)+7 \kappa _{\rho }^2 \kappa _{\tau }^4-5 \kappa _{\rho }^6\bigg),
\\[2ex]
    \mathcal{R}_{0T\perp T}&=\frac{128 \sqrt{2} }{\sqrt{\varepsilon _{\rho }^2-\kappa _{\rho }^2}}
    \bigg(\kappa _{\rho }^4 \left(2 \varepsilon _{\rho }+\kappa _{\tau }^2-2\right)+\varepsilon _{\rho } \left(2 \varepsilon _{\rho }-1\right) \kappa _{\tau }^4\nonumber\\
    &+\kappa _{\rho }^2 \kappa _{\tau }^2 \Big(\left(\varepsilon _{\rho }-2\right) \kappa _{\tau }^2-4 \varepsilon _{\rho }^2+2 \varepsilon _{\rho }+1\Big)\bigg),
\\[2ex]
    \mathcal{R}_{SPt}&=-32 \kappa _{\tau } \left(-2 \varepsilon _{\rho } \kappa _{\tau }^2+\left(4 \varepsilon _{\rho }-1\right) \kappa _{\rho }^2-2 \kappa _{\rho }^4+\kappa _{\tau }^4\right),
\\[2ex]
    \mathcal{R}_{SP0T}&=2 \sqrt{2}\mathcal{R}_{SP\parallel T}=\frac{1}{\kappa _{\tau }}\mathcal{R}_{t0T}=\frac{2 \sqrt{2}}{\kappa _{\tau }}\mathcal{R}_{t\parallel T}=\frac{128}{\sqrt{\varepsilon _{\rho }^2-\kappa _{\rho }^2}} \bigg(\kappa _{\tau }^4 \left(\kappa _{\rho }^2-\varepsilon _{\rho }\right)\nonumber\\
    &+2 \kappa _{\tau }^2 \Big(\left(\varepsilon _{\rho }-1\right) \kappa _{\rho }^2+\varepsilon _{\rho }^2-\kappa _{\rho }^4\Big)+\varepsilon _{\rho } \kappa _{\rho }^2 \left(-4 \varepsilon _{\rho }+2 \kappa _{\rho }^2+1\right)+\kappa _{\rho }^4\bigg),
\\[2ex]
    \mathcal{R}_{00 T,1}&=\frac{32}{\kappa _{\tau } \left(\varepsilon _{\rho }^2-\kappa _{\rho }^2\right)} \bigg(4 \varepsilon _{\rho }^3 \kappa _{\tau }^2 \left(\kappa _{\tau }^2-2 \kappa _{\rho }^2\right)+4 \varepsilon _{\rho }^2 \Big(\left(\kappa _{\rho }^2-1\right) \kappa _{\tau }^4+2 \kappa _{\rho }^4\Big)\nonumber\\
    &-\varepsilon _{\rho } \left(\kappa _{\rho }^2+1\right) \left(2 \kappa _{\rho }^2-\kappa _{\tau }^2\right) \left(\kappa _{\rho }^2+\kappa _{\tau }^4\right)-2 \kappa _{\rho }^2 \left(\kappa _{\tau }^2-1\right) \left(\kappa _{\rho }^2 \kappa _{\tau }^2-2 \kappa _{\rho }^4+\kappa _{\tau }^4\right)\bigg),
\\[2ex]
     \mathcal{R}_{00 T,2}&=2 \sqrt{2}\mathcal{R}_{0\parallel T}=2 \sqrt{2}\mathcal{R}_{\parallel 0T}=-4\mathcal{R}_{\parallel\parallel T,2}=-4\mathcal{R}_{\perp\perp T,2}\nonumber\\
     &=\frac{32}{\kappa _{\tau } \left(\varepsilon _{\rho }^2-\kappa _{\rho }^2\right)} \left(2 \kappa _{\rho }^2-\kappa _{\tau }^2\right) \bigg(-4 \varepsilon _{\rho }^3 \kappa _{\tau }^2+4 \varepsilon _{\rho }^2 \left(\kappa _{\tau }^2+1\right) \left(\kappa _{\rho }^2+\kappa _{\tau }^2\right)\nonumber\\
     &-\varepsilon _{\rho } \Big(\kappa _{\rho }^2 \left(3 \kappa _{\tau }^4+8 \kappa _{\tau }^2+3\right)+3 \kappa _{\rho }^4+3 \kappa _{\tau }^4\Big)+2 \kappa _{\rho }^2 \left(\kappa _{\tau }^2+1\right) \left(\kappa _{\rho }^2+\kappa _{\tau }^2\right)\bigg),
\\[2ex]
    \mathcal{R}_{0\perp T}&=-\frac{1}{\sqrt{2}}\mathcal{R}_{\parallel\perp T}=\mathcal{R}_{\perp 0T}=-\frac{1}{\sqrt{2}}\mathcal{R}_{\perp\parallel T}\nonumber\\
    &=\frac{16 \sqrt{2}}{\kappa _{\tau } \sqrt{\varepsilon _{\rho }^2-\kappa _{\rho }^2}}
    \bigg(-\kappa _{\rho }^4 \Big(\left(1-4 \varepsilon _{\rho }\right) \kappa _{\tau }^2+2 \kappa _{\tau }^4-2\Big)\nonumber\\
    &+\kappa _{\rho }^2 \kappa _{\tau }^2 \left(-2 \varepsilon _{\rho }+\kappa _{\tau }^4+2 \kappa _{\tau }^2-1\right)+\left(1-2 \varepsilon _{\rho }\right) \kappa _{\tau }^6-2 \kappa _{\rho }^6\bigg),
\\[2ex]
\label{RparparT1}
    \mathcal{R}_{\parallel\parallel T,1}&=\mathcal{R}_{\perp\perp T,1}\nonumber\\
    &=\frac{8}{\kappa _{\tau } \left(\varepsilon _{\rho }^2-\kappa _{\rho }^2\right)}
    \bigg(4 \varepsilon _{\rho }^3 \kappa _{\tau }^2 \left(\kappa _{\tau }^2-2 \kappa _{\rho }^2\right)+4 \varepsilon _{\rho }^2 \left(\kappa _{\tau }^2-1\right) \left(\kappa _{\rho }^2 \kappa _{\tau }^2-2 \kappa _{\rho }^4+\kappa _{\tau }^4\right)\nonumber\\
    &+\varepsilon _{\rho } \left(2 \kappa _{\rho }^2-\kappa _{\tau }^2\right) \Big(\kappa _{\rho }^2 \left(\kappa _{\tau }^4+8 \kappa _{\tau }^2+1\right)+\kappa _{\rho }^4+\kappa _{\tau }^4\Big)+4 \kappa _{\rho }^6 \left(\kappa _{\tau }^2-3\right)\nl
    &+2 \kappa _{\rho }^4 \left(\kappa _{\tau }^2-3 \kappa _{\tau }^4\right)-2 \kappa _{\rho }^2 \kappa _{\tau }^4 \left(\kappa _{\tau }^2-3\right)\bigg),
\end{align}
The coefficients in Table~\ref{imaginarypart of the cross term} are
\begin{align}\label{Itper}
    \mathcal{I}_{t\perp}&=\kappa _{\tau }\mathcal{I}_{SP\perp}=\frac{8 \sqrt{2} }{\sqrt{\varepsilon _{\rho }^2-\kappa _{\rho }^2}} \bigg(-2 \kappa _{\rho }^4 \Big(\varepsilon _{\rho } \left(\kappa _{\tau }^2+2\right)-1\Big)\nonumber\\
    &+\Big(\varepsilon _{\rho } \left(4 \varepsilon _{\rho }-1\right)-1\Big) \kappa _{\rho }^2 \kappa _{\tau }^2+\left(\varepsilon _{\rho }-1\right) \kappa _{\tau }^4 \left(\kappa _{\tau }^2-2 \varepsilon _{\rho }\right)+2 \kappa _{\rho }^6\bigg),
    \\[2ex]
    \mathcal{I}_{0\perp}&=\frac{1}{\sqrt{2}}\mathcal{I}_{\parallel\perp}=\frac{2 \sqrt{2}}{\kappa _{\tau }^2 \left(\varepsilon _{\rho }^2-\kappa _{\rho }^2\right)}
    \bigg(-4 \varepsilon _{\rho }^3 \kappa _{\tau }^2 \left(\kappa _{\tau }^2+1\right) \left(\kappa _{\tau }^2-2 \kappa _{\rho }^2\right)\nonumber\\
    &-2 \varepsilon _{\rho } \left(\kappa _{\tau }^2+1\right) \left(-2 \kappa _{\rho }^4 \kappa _{\tau }^2+\kappa _{\rho }^2 \kappa _{\tau }^4-6 \kappa _{\rho }^6+3 \kappa _{\tau }^6\right)\nonumber\\
    &-2 \varepsilon _{\rho }^2 \Big(2 \kappa _{\rho }^4 \left(\kappa _{\tau }^4+6 \kappa _{\tau }^2+1\right)+2 \kappa _{\rho }^2 \kappa _{\tau }^4-\kappa _{\tau }^4 \left(\kappa _{\tau }^4+6 \kappa _{\tau }^2+1\right)\Big)\nonumber\\
    &-2 \kappa _{\rho }^6 \left(\kappa _{\tau }^4+1\right)+\kappa _{\rho }^4 \kappa _{\tau }^4+\kappa _{\rho }^2 \left(\kappa _{\tau }^8+\kappa _{\tau }^4\right)-6 \kappa _{\rho }^8+3 \kappa _{\tau }^8\bigg),
   \\[2ex]
    \mathcal{I}_{SP\perp T}&=\frac{1}{\kappa _{\tau }}\mathcal{I}_{t\perp T}=\frac{32 \sqrt{2} }{\sqrt{ \varepsilon _{\rho }^2-\kappa _{\rho }^2}}
    \bigg(\varepsilon _{\rho }^2 \left(4 \kappa _{\rho }^2-2 \kappa _{\tau }^2\right)\nonumber\\
    &-\varepsilon _{\rho } \Big(\kappa _{\rho }^2 \left(2 \kappa _{\tau }^2+1\right)+2 \kappa _{\rho }^4-\kappa _{\tau }^4\Big)+\kappa _{\rho }^4 \left(2 \kappa _{\tau }^2-1\right)-\kappa _{\rho }^2 \kappa _{\tau }^2 \left(\kappa _{\tau }^2-2\right)\bigg),
    \\[2ex]
    \mathcal{I}_{0\parallel T}&=-\mathcal{I}_{\parallel0 T}=\frac{16 \sqrt{2} }{\kappa _{\tau } \sqrt{\varepsilon _{\rho }^2-\kappa _{\rho }^2}}
     \bigg(\kappa _{\rho }^4 \Big(\left(1-4 \varepsilon _{\rho }\right) \kappa _{\tau }^2+2 \kappa _{\tau }^4-2\Big)\nonumber\\
     &-\kappa _{\rho }^2 \kappa _{\tau }^2 \left(-2 \varepsilon _{\rho }+\kappa _{\tau }^4+2 \kappa _{\tau }^2-1\right)+\left(2 \varepsilon _{\rho }-1\right) \kappa _{\tau }^6+2 \kappa _{\rho }^6\bigg),
  \\[2ex]
   \label{I0perT}
    \mathcal{I}_{0\perp T}&=\frac{1}{\sqrt{2}}\mathcal{I}_{\parallel\perp T}=-\mathcal{I}_{\perp 0T }=-\frac{1}{\sqrt{2}}\mathcal{I}_{\perp\parallel T}\nonumber\\
    &=-\frac{8 \sqrt{2}}{\kappa _{\tau } \left(\varepsilon _{\rho }^2-\kappa _{\rho }^2\right)}
    \left(2 \kappa _{\rho }^2-\kappa _{\tau }^2\right)
    \bigg(-4 \varepsilon _{\rho }^3 \kappa _{\tau }^2+4 \varepsilon _{\rho }^2 \left(\kappa _{\tau }^2+1\right) \left(\kappa _{\rho }^2+\kappa _{\tau }^2\right)\nonumber\\
    &-\varepsilon _{\rho } \Big(\kappa _{\rho }^2 \left(3 \kappa _{\tau }^4+8 \kappa _{\tau }^2+3\right)+3 \kappa _{\rho }^4+3 \kappa _{\tau }^4\Big)+2 \kappa _{\rho }^2 \left(\kappa _{\tau }^2+1\right) \left(\kappa _{\rho }^2+\kappa _{\tau }^2\right)\bigg).
\end{align}

\section{Expressions of $\bar Y_h(q^2,E_\rho)$ and $\Delta Y_h(q^2,E_\rho)$}
\label{YY}

\begin{align}
    \bar{Y}_h(q^2,E_\rho)&=\frac{1}{2~\Gamma(\tau\rightarrow\rho\nu_\tau)}\left(\frac{d\Gamma(\tau_\uparrow\rightarrow\rho_h\nu_\tau)}{dE_\rho}+\frac{d\Gamma(\tau_\downarrow\rightarrow\rho_h\nu_\tau)}{dE_\rho}\right),\\
    \Delta{Y}_h(q^2,E_\rho)&=\frac{1}{2~\Gamma(\tau\rightarrow\rho\nu_\tau)}\left(\frac{d\Gamma(\tau_\uparrow\rightarrow\rho_h\nu_\tau)}{dE_\rho}-\frac{d\Gamma(\tau_\downarrow\rightarrow\rho_h\nu_\tau)}{dE_\rho}\right),
\end{align}
where in the $\tau$ rest frame the $\tau\to \rho\nu_\tau$ width reads
\begin{align}    \Gamma(\tau\rightarrow\rho\nu_\tau)=\frac{G_F^2|V_{ud}|^2f_\rho^2m_\tau^3}{16\pi}\left(1-\frac{m_\rho^2}{m_\tau^2}\right)^2\left(1+\frac{2m_\rho^2}{m_\tau^2}\right),
\end{align}
and in $W^*$ rest frame
\begin{align}
\frac{d\Gamma(\tau_\uparrow\rightarrow\rho_L\nu_\tau)}{dE_\rho}&=
\frac{ G_F^2 |V_{ud}|^2 f_{\rho }^2~ m_{\tau } \left(2E_\rho m_\tau^2\sqrt{q^2}-m_\rho^2q^2-m_\tau^4\right) }{4 \pi  \sqrt{q^2} \left(E_{\rho }^2-m_{\rho }^2\right) \left(m_{\tau }^2-q^2\right){}^2}\left(m_{\rho }^2-E_{\rho } \sqrt{q^2}\right){}^2,\\
\frac{d\Gamma(\tau_\downarrow\rightarrow\rho_L\nu_\tau)}{dE_\rho}&=
    \frac{ G_F^2 |V_{ud}|^2 f_{\rho }^2 \sqrt{q^2}\left(m_{\rho }^2-2 E_{\rho } \sqrt{q^2}+q^2\right) }{4 \pi~ m_{\tau } \left(E_{\rho }^2-m_{\rho }^2\right) \left(m_{\tau }^2-q^2\right){}^2}\left(m_{\tau }^2E_{\rho } - m_{\rho }^2\sqrt{q^2}\right){}^2,\\ \frac{d\Gamma\left(\tau_\uparrow\rightarrow\rho_T\nu_\tau\right)}{dE_\rho}&=
    \frac{G_F^2 |V_{ud}|^2f_\rho^2~ m_\rho^2~ m_\tau\left(m_\rho^2-2E_\rho\sqrt{q^2}+q^2\right)}{4\pi\sqrt{q^2}(E_\rho^2-m_\rho^2)\left(m_\tau^2-q^2\right)^2}\nonumber\\
 &\times\left(m_\tau^4-2m_\tau^2E_\rho \sqrt{q^2}+\left(2E_\rho^2-m_\rho^2\right)q^2\right),\\
\frac{d\Gamma\left(\tau_\downarrow\rightarrow\rho_T\nu_\tau\right)}{dE_\rho}&=
    \frac{G_F^2 |V_{ud}|^2f_\rho^2~ m_\rho^2 \sqrt{q^2}\left(2m_\tau^2E_\rho \sqrt{q^2}-m_\rho^2q^2-m_\tau^4\right)}{4\pi~m_\tau(E_\rho^2-m_\rho^2)\left(m_\tau^2-q^2\right)^2}\nonumber\\
    &\times\left(2E_\rho^2-2E_\rho\sqrt{q^2}+q^2-m_\rho^2\right).
\end{align}
\end{appendix}

\end{document}